\documentclass[12pt]{article}
\usepackage[dvips]{epsfig}
\usepackage{latexsym}
\usepackage[bf,footnotesize]{caption}	
\begin{document}
\setlength{\unitlength}{1mm}


\newcommand{\ket}[1] {\mbox{$ \vert #1 \rangle $}}
\newcommand{\bra}[1] {\mbox{$ \langle #1 \vert $}}
\def\vac{\ket{0}} 
\def\thermal{\ket{\beta}}
\def\bvac{\bra{0}}
\def\bthermal{\bra{\beta}}
\newcommand{\ave}[1] {\mbox{$ \langle #1 \rangle $}}
\newcommand{\vacave}[1] {\mbox{$ \bvac #1 \vac $}}
\newcommand{\thermalave}[1] {\mbox{$ \bthermal #1 \thermal $}}
\newcommand{\scal}[2]{\mbox{$ \langle #1 \vert #2 \rangle $}}
\newcommand{\expect}[3] {\mbox{$ \bra{#1} #2 \ket{#3} $}}
\newcommand{\ki}{\mbox{$ \ket{\psi_i} $}}
\newcommand{\bi}{\mbox{$ \bra{\psi_i} $}}
\def\a{\hat{a}}
\def\b{\hat{b}}


\def\t{\tau}
\def\ga{\gamma}
\def\Ga{\Gamma}
\def\om{\omega}
\def\omp{\om^\p}
\def\Om{\Omega}
\def\la{\lambda}
\def\lap{\lambda^\p}
\def\mup{\mu^\p}
\def\lp{l^\p}
\def\kp{k^\p}
\def\sig{\sigma} 
\def\al{\alpha}
\def\bt{\beta}
\def\alb{\bar\alpha}
\def\btb{\bar\beta}
\def\e{\epsilon}
\def\psip{\stackrel{.}{\psi}}
\def\fp{\stackrel{.}{f}}
\def\VB{{\bar V}}
\def\UB{{\bar U}}
\def\vb{{\bar v}}
\def\ub{{\bar u}}
\def\ffi{\varphi}
\def\scryp{{\cal J}^+}
\def\scrym{{\cal J}^-}
\def\p{\prime}
\def\Phic{\Phi^\dagger}
\def\Phiv{\Phi^V}
\def\Phivc{{\Phi^V}^\dagger}
\def\Phivp{\Phi^{V'}}
\def\Phivpc{{\Phi^{V'}}^\dagger}
\def\Phivb{\Phi^{\VB}}
\def\Phivbc{{\Phi^{\VB}}^\dagger}
\def\Phiu{\Phi^U}
\def\Phiuc{{\Phi^U}^\dagger}
\def\Phiup{\Phi^{U'}}
\def\Phiupc{{\Phi^{U'}}^\dagger}
\def\Phiub{\Phi^{\UB}}
\def\Phiubc{{\Phi^{\UB}}^\dagger}

\def\dw{\partial_\omega}
\def\dla{\partial_\la}
\def\dt{\partial_t}
\def\dtt{\partial^{2}_t}
\def\dz{\partial_z}
\def\dzz{\partial^{2}_z}
\def\dU{\partial_U}
\def\div{\partial_V}
\def\divp{\partial_{V'}}
\def\divb{\partial_{\VB}}
\def\di{\partial}
\def\diditau{\raise 0.1mm \hbox{$\stackrel{\leftrightarrow}{\di_\tau}$}}
\def\didiv{\raise 0.1mm \hbox{$\stackrel{\leftrightarrow}{\di_V}$}}
\def\didiu{\raise 0.1mm \hbox{$\stackrel{\leftrightarrow}{\di_U}$}}


\def\U{{\cal U}}
\def\V{{\cal V}}
\def\B{{\cal B}}
\def\L{\tilde L}
\def\LR{{\cal L}_{int}^R}
\def\LL{{\cal L}_{int}^L}
\def\JR{J^R_{int}}
\def\JL{J^L_{int}}
\def\gr{g_R}
\def\gl{g_L}
\def\g{\tilde g}
\def\TVV{T_{VV}}
\def\TVVB{T_{{\VB}{\VB}}}
\def\Tvv{T_{vv}}
\def\Tvvb{T_{{\vb}{\vb}}}
\newcommand{\im} {\Im m}
\newcommand{\re} {\Re e}
\def\S{{\mathbf S}}
\def\T{{\mathbf T}}
\def\1{{\mathbf 1}}
\def\Ss{\mbox{$\hat S$}}

\def\disp{\displaystyle}
\def\bitem{\begin{itemize}}
\def\eitem{\end{itemize}}
\def\bes{\begin{description}}
\def\es{\end{description}}
\newcommand{\be} {\begin{equation}}
\newcommand{\ee} {\end{equation}}
\newcommand{\ba} {\begin{eqnarray}}
\newcommand{\ea} {\end{eqnarray}}

\def\cf{{\it cf}~}
\def\ie{{\it i.e.}~}
\def\etc{{\it etc}...}
\def\where{\mbox{where} \; \;}
\def\with{\mbox{with} \; \;}
\def\for{\mbox{for} \; \;}
\def\and{\mbox{and} \; \;}
\def\eg{{\it e.g.}~}

\def\nn{\nonumber \\}
\newcommand{\reff}[1]{eq. (\ref{#1})}

\def\dterm{${\cal D}$-term ~}
\def\dpart{${\cal D}$-part ~}


\def\k{\tilde{k}}
\def\r{\Big]}
\def\l{\Big[}
\def\sgn{\mbox{sgn}}
\def\th{\mbox{th}}
\def\ch{\mbox{ch}}
\def\sh{\mbox{sh}}
\def\d4{d^4x}
\def\half{{1 \over 2}}

\def\inte{\int_{-\infty}^{+\infty}}
\def\into{\int_{0}^{\infty}}
\def\intetau{\inte \! d\tau \;}
\def\intet{\inte \! dt \;}                        
\def\intetp{\inte \! dt' \;}
\def\interho{\inte \! d\rho \;}
\def\intetaup{\inte \! d\tau' \;}
\def\interhop{\inte \! d\rho' \;}
\def\intev{\inte \! dV \;}
\def\intevv{\inte \! dv \;}
\def\intevb{\inte \! d\VB \;}
\def\intevp{\inte \! dV' \;}
\def\intemu{\inte \! d\mu \;}
\def\intemup{\inte \! d\mup \;}
\def\inteom{\inte \! d\om \;}
\def\inteomp{\inte \! d\omp \;}
\def\intela{\inte \! d\la \;}
\def\intelap{\inte \! d\lap \;}
\def\intel{\inte \! dl \;}
\def\intelp{\inte \! dl^\p \;}
\def\intek{\inte \! dk \;}
\def\intekp{\inte \! dk^\p \;}

\def\intotau{\into \! d\tau \;}
\def\intot{\into \! dt \;}
\def\intotp{\into \! dt' \;}
\def\intorho{\into \! d\rho \;}
\def\intotaup{\into \! d\tau' \;}
\def\intorhop{\into \! d\rho' \;}
\def\intov{\into \! dV \;}
\def\intovp{\into \! dV' \;}
\def\intomu{\into \! d\mu \;}
\def\intomup{\into \! d\mup \;}
\def\intoom{\into \! d\om \;}
\def\intoomp{\into \! d\omp \;}
\def\intola{\into \! d\la \;}
\def\intolap{\into \! d\lap \;}
\def\intol{\into \! dl \;}
\def\intolp{\into \! dl^\p \;}
\def\intok{\into \! dk \;}
\def\intokp{\into \! dk^\p \;}


\def\M{Minkowski}
\def\sg{\sqrt{-g}}
\def\sga{\sqrt{-\gamma}}
\def\gmn{g_{\mu\nu}}
\def\gnm{g^{\mu\nu}}
\def\gmn{g_{\mu\nu}}
\def\Rmn{R_{\mu\nu}}
\def\Rnm{R^{\mu\nu}}
\def\Gmn{G_{\mu\nu}}
\def\Gnm{G^{\mu\nu}}
\def\Tmn{T_{\mu\nu}}
\def\Tnm{T^{\mu\nu}}
\def\gnm{g^{\mu\nu}}
\def\Rij{R_{ij}}
\def\Rji{R^{ij}}
\def\Kij{K_{ij}}
\def\Kji{K^{ij}}
\def\gij{g_{ij}}
\def\gji{g^{ij}}
\def\piij{\pi_{ij}}
\def\piji{\pi^{ij}}
\def\gaij{\gamma_{ij}}
\def\gaji{\gamma^{ij}}
\def\H{{\cal H}}
\def\Hi{{\cal H}_i}
\def\Hii{{\cal H}^i}
\newcommand{\G}[3]{\Gamma^{#1}_{#2 \: #3}}
\def\A{{\cal A}_{BH}}

\overfullrule=0pt \def\sqr#1#2{{\vcenter{\vbox{\hrule height.#2pt
          \hbox{\vrule width.#2pt height#1pt \kern#1pt
           \vrule width.#2pt}
           \hrule height.#2pt}}}}
\def\square{\mathchoice\sqr68\sqr68\sqr{4.2}6\sqr{3}6} \def\lrpartial{\mathrel
{\partial\kern-.75em\raise1.75ex\hbox{$\leftrightarrow$}}}

\begin{flushright}
\end{flushright}
\vskip 1. truecm
\vskip 1. truecm
\centerline{\Large\bf{Notes on moving mirrors}}
\vskip 1. truecm
\vskip 1. truecm

\centerline{{\bf N. Obadia}\footnote{e-mail: obadia@celfi.phys.univ-tours.fr}
 and {\bf
R. Parentani}\footnote{e-mail: parenta@celfi.phys.univ-tours.fr}}
\vskip 5 truemm\vskip 5 truemm
\centerline{Laboratoire de Math\'ematiques et Physique Th\'eorique,
CNRS-UMR 6083}
\centerline{Parc de Grandmont, 37200 Tours, France.}
\vskip 10 truemm

\vskip 1.5 truecm

\vskip 1.5 truecm
\centerline{{\bf Abstract }}
\vskip 3 truemm
\noindent
The Davies-Fulling (DF) model
describes the scattering of a massless field
by a non-inertial mirror in two dimensions. 
In this paper, we generalize this model in two different ways.
First, we consider partially reflecting mirrors. 
We show that the Bogoliubov coefficients 
relating inertial modes can be expressed 
in terms of the frequency dependent reflection factor which is 
specified in the rest frame of the mirror
and the transformation from the inertial modes to the 
modes at rest with respect to the mirror.
In this perspective,  
the DF model is simply the limiting case when
this factor is unity for all frequencies.
In the second part, we introduce an alternative model
which is based on self-interactions described by an action 
principle. When the coupling is constant, this model can be 
solved exactly and gives rise to a partially reflecting mirror.
The usefulness of this dynamical model lies in the possibility 
of switching off the coupling between the mirror and the field. 
This allows to obtain regularized expressions
for the fluxes in situations where they are singular 
when using the DF model. Two examples are considered. 
The first concerns the flux induced by the disappearance
of the reflection condition, a situation which bears 
some analogies with the end of the evaporation of a black hole.
The second case concerns the flux emitted 
by a uniformly accelerated mirror.

\vfill
\newpage


\section{Introduction}

\par
The Davies-Fulling (DF) model \cite{DF}
describes the scattering of a massless field 
by a non-inertial mirror in two dimensions. 
It has received and continues to receive attention 
\cite{BD,Grove,Carlitz,Wilczek,ChungVerlinde,Strominger,
Primer,Ford,recmir,P:hawkfeynman}
principally because of its simplicity 
and its relationship with Hawking radiation \cite{Hawking}. 
(One can indeed mimic the varying Doppler effect  
induced by the collapse of a star
by the total reflection on a receding mirror.) 
Because of its simplicity, this model has been also used to 
investigate the relationships between the particle description 
of fluxes emitted by the mirror and its field description 
based on the two-point Green function. 
The motivation behind this analysis is the following.
When quantizing a field in a curved space-time
one looses the uniqueness of choice 
for the particle notion which is then used 
to define the vacuum and its excitations. 
Based on this fact, some authors have proposed 
to discard the particle point of view\cite{Waldbook}.
The DF model, being defined in flat space time 
and giving rise to particle creation as in a curved space-time, 
provides a good playground for confronting the two points of view.
Finally, the DF model also provides a good starting point 
for studying the role of ultra-high frequencies which 
arise in the presence of event horizons \cite{Stephens,Englert,MaPa,Kiem}. 
This is particularly true 
when considering uniformly accelerated 
mirrors\cite{Grove,Gerlach,FrolovSingh}. 
Indeed, in this case one has to confront the fact that
the instantaneous value of the energy flux identically vanishes 
whereas the Bogoliubov coefficients, 
mixing positive and negative frequencies, do not vanish 
and lead to a total energy which furthermore diverges.

Quite independently of these specific difficulties,
there is a fundamental reason which renders the analysis 
of these problems complicated: 
the DF model does not follow from an action principle.
In fact, the reflection condition is imposed from the outset 
instead of following from interactions with the boundary. 
Therefore only questions concerning asymptotic properties 
of asymptotically inertial mirrors 
can be properly answered. 
To emphasize this point, we shall show in the first part of this article
that the scattering in the DF model 
can be expressed in purely kinematic terms. 
It results from the Bogoliubov transformation 
relating the usual Minkowski modes to non inertial modes 
which are eigen-modes with respect to the proper time of the mirror. 
The scattering of the latter is then trivial, 
as trivial as the scattering of Minkowski modes 
by an inertial mirror.
This rephrasing of the DF model is very useful 
in that it allows to consider partially transmitting mirrors
with arbitrary frequency dependent transmission coefficients. 
In this perspective, the DF model is simply the limiting case
in which the reflection is total for all frequencies.  

In the second part of the paper
we analyze an alternative model
for the scattering along a given trajectory 
which is based on self-interactions described by an action principle.
The main motivation for considering this model
is that we can now switch on and off the coupling between 
the mirror and the field.
Therefore, we can work with well defined asymptotic free states.
The relationship between the partially transmitting mirrors
previously considered and this model based on asymptotic states
will be explicitly made.

In the first part of this Section
we work with a coupling which is constant.
In this case, the  Born series can be exactly summed 
and lead to a partially transmitting mirror.
Moreover, in the large coupling constant limit,
one recovers the DF model, i.e. total reflection.
Secondly, we consider the case when the coupling 
is time dependent. 
In this case, we compute the fluxes perturbatively
to quadratic order in the coupling.
The novelty arises from transients
effects associated with the switching on and off.
The possibility of controlling these
transients is crucial for regularizing the fluxes
in situations where they are singular when using the DF model.

To make this explicit, we consider two examples.
The first one consists in computing the flux associated with 
the appearance (or disappearance) of the reflecting boundary condition.
This problem was considered by Anderson and deWitt \cite{AndersonDeWitt}.
Moreover, as discussed in \cite{Strominger},
it presents 
some analogies with the residual flux associated with the
disappearance of a black hole at the end of the 
evaporation process.
When using the DF model, the flux is singular and 
its spectral properties are ill defined. 
On the contrary, with the self-interacting model,
it can be described by
a well-defined process in which the 
coupling to the mirror is switched off more and more rapidly.
The second application concerns the flux emitted by a uniformly
accelerated mirror. In the DF model, the energy flux vanishes everywhere
but on the horizons where it is not defined. 
With the other model instead, a well-defined and 
regular expression will be obtained.
In the intermediate regime we recover the vanishing of the local flux. 
But we also find transient effects which become singular 
when the switching on and off of 
the coupling is performed for asymptotic early and late proper times.

Finally, we wish to stress that in this paper
the recoil effects shall be totally ignored 
since the trajectory of the mirror 
is given once for all.
Nevertheless, since the interacting model is based on Feynman diagrams, 
it prepares the analysis of taking into account 
the dynamics of the mirror\cite{recmir,P:hawkfeynman}.
Indeed the $S$-matrices computed 
with or without back-reaction effects
possess a very similar structure. 

\section{The kinematic models}
\par
In the first part of this Section, we review 
the basic properties of the Davies-Fulling model.
In particular we compare the particle description of the fluxes 
based on Bogoliubov coefficients with that based on the
two-point functions.
In the second part,
we show how the scattering process can be generalized 
so as to describe partially transmitting mirrors.
This generalization will be performed in a matrix formalism. 
We have chosen this formalism for two reasons:
first to emphasize the kinematic nature of the DF model,
and secondly to introduce in natural terms 
the generalization to partial reflection.
In the third part, we relate the Bogoliubov coefficients
to the $S$-matrix acting in the Fock space, thereby preparing
the analysis of transition amplitudes 
performed in the next Section.

\subsection{The Davies-Fulling model}

\par
In the Davies-Fulling model, 
the mirror is perfectly reflecting for all frequencies
and its trajectory is chosen from the outset. 
Moreover, no width is attributed to the reflecting condition, 
i.e. it acts like a delta in space.
Beside the fact that the trajectory is always time-like, 
we shall also impose that it is asymptotically inertial. 
In conformal terms this means 
that the trajectory starts from $i^-$ and ends in $i^+$, 
the past and future time-like infinities 
respectively \cite{MTW}. 
The reason is that in the other cases, 
i.e. when the mirror originates and/or ends on null infinities, 
the calculation of the energy radiated
by the mirror is ill-defined.
(The specific problems associated with such asymptotic 
trajectories will be considered in a next article \cite{next}).

In this paper,
we shall consider the scattering of a {\it complex} massless scalar field.  
The reason for this choice is that it allows to consider more general 
scattering matrices when the reflection condition is not perfect. 
This possibility will be exploited in the next subsections.
Since the dynamics of the mirror is trivial,
the evolution of the field is entirely governed 
by the d'Alembert equation 
\ba \label{1}
(\partial_t^2 -\partial_z^2 )  
\Phi(t,z) 
= 0
\ea
and the reflection condition
\ba \label{2}
\Phi(t,z_{cl}(t))=0
\ea
along the classical trajectory of the mirror $z=z_{cl}(t)$.

Since the field is massless and since we work in two dimensions, 
it is particularly useful to work in the light-like coordinates 
defined by $U, V = t \mp z$.
For instance, 
the general solution of \reff{1} is the sum 
of a function of $U$ alone plus a function of $V$.
Moreover, since the mirror is perfectly reflecting, 
the trajectory of the mirror completely decouples 
the left hand side configurations from the right hand side ones. 
Therefore, in this subsection, we can and shall restrict 
our attention to the configurations living on the left of the mirror.

Finally, 
since the mirror trajectory emerges from $i^-$, 
$V=-\infty$ is a complete Cauchy surface. 
Hence, the usual modes of the d'Alembertian given by
\ba \label{usualmode}
\ffi_k(U) 
= \frac{e^{-ik U}}{\sqrt{4\pi|k|}}
\ea
form a complete and orthonormal basis.
(Instead, when the trajectory starts from the left part of $\scrym$, 
the choice of a complete and orthonormal basis should be 
reconsidered\cite{next}.)
We recall that the norm of the modes is determined by the
Klein-Gordon scalar product which reads, 
when evaluated on $\scrym$ :
\be \label{normalization}
\langle \ffi_k \vert \ffi_{k'} \rangle 
= \inte dU  \ffi_k^* i \didiu  \ffi_{k'}
= sgn (k) \: \delta(k - k') \ . 
\ee
The scattered mode corresponding to \reff{usualmode} 
is determined by \reff{2} to be
\ba
\ffi_k^{scat}(V) 
= -\frac{e^{-ik U_{cl}(V)}}{\sqrt{4\pi|k|}} \ ,
\label{scatt}
\ea
where $U=U_{cl}(V)$ is the trajectory of the mirror in the light-like
coordinates.

The $in$-mode $\ffi_k^{in}(U, V)$ is by definition
the solution of eqs. (\ref{1}) and (\ref{2}) 
which has \reff{usualmode} as initial data. It is given by
\ba \label{inmodes}
\ffi_k^{in}(U, V) 
= \frac{e^{-ik U}}{\sqrt{4\pi|k|}} 
  - \frac{e^{-ik U_{cl}(V)}}{\sqrt{4\pi|k|}} \ .
\ea 
To analyze the frequency content of its scattered part,
it should be Fourier decomposed 
on the final Cauchy surface $U= + \infty$ (the left part of $\scryp$). 
In total analogy with what we have on $\scrym$, on  $\scryp$
the normalized modes are 
\ba \label{outmod}
\ffi_\om(V) 
= \frac{e^{-i\om V}} {\sqrt{4\pi|\om|}} \ .
\ea
Then the scattered mode (\ref{scatt}) can be decomposed as  
\ba \label{decomp}
\ffi_k^{scat}
= \intoom \Big( 
\al_{\om k}^* \; \ffi_\om - \bt_{\om k}^* \; \ffi_\om^* 
\Big)
\ea
where the coefficients $ \al_{\om k},  \bt_{\om k} $ 
are given by the overlaps
\ba \label{overlap}
\al_{\om k}^*
&=& 
\scal{\ffi_\om}{\ffi^{scat}_{k}} 
= 
-2 \: \intev \frac{e^{i \om V}}{\sqrt{4 \pi |\om|^{-1}}} 
   \: \frac{e^{-i k U_{cl}(V)}}{\sqrt{4 \pi |k|}}
\nn
\bt_{\om k}^*
&=& 
\scal{\ffi_\om^*}{\ffi^{scat}_{k}} \ .
\ea
Since both the initial and the final set of modes are complete, 
the coefficients $ \al_{\om k},  \bt_{\om k} $ satisfy the relations
\ba \label{bogcoef}
\intok ( 
\al_{\om k}^*  \al_{\omp k}  - \bt_{\om k}  \bt^*_{\omp k} 
) 
&=& \delta ( \om - \omp) 
\nn
\intoom ( 
\al_{\om k}  \al^*_{\om k'}  - \bt_{\om k}  \bt^*_{\om k'} 
) 
&=& \delta ( k - k') 
\nn
\intok ( 
\al_{\om k} \bt_{\om k'} - \bt_{\om k} \al_{\om k'}) &=& 0
\nn
\intoom ( 
\al_{\om k} \bt^*_{\omp k} - \bt^*_{\om k} \al_{\omp k}) &=& 0 \ .
\ea
Note that these relations are not trivially fulfilled 
when the trajectory of the mirror reaches 
one of the null infinities rather than the time-like ones.
Notice also that the overlaps (\ref{overlap})
can be computed on any space-like surface 
which runs from $z=-\infty$ to some point on the mirror
$(t, z_{cl}(t))$.
In this case, one should use the full expression 
of the $in$ modes given in \reff{inmodes}
as well as that of the $out$ modes given by
\be \label{outmodes}
\varphi_{\om}^{out} (U, V) 
=  \varphi_\om (V) +  \varphi^{bscat}_{\om}(U) \ .
\ee
The second term $\varphi^{bscat}_{\om}$
results from the backward scattering 
of $\ffi_\om$ given in \reff{outmod}.

When the overlaps $ \al_{\om k}$  and $ \bt_{\om k} $ are known,
the classical scattering problem is solved.
That is, it suffices to decompose the initial data 
in terms of the modes (\ref{outmod}) to obtain, 
through \reff{overlap}, 
the Fourier content of its image on $\scryp$. 
It should be pointed out 
that the coefficients $\bt_{\om k}$ 
which mix positive and negative frequencies 
have a well defined role in this classical wave theory: 
they determine the (non-adiabatic \cite{wdwpc}) 
increase of the Fourier components 
of the scattered wave 
(see \eg eq. (11) in \cite{recmir}
for their influence on the energy of the reflected wave).
It should be also pointed out 
that one can recover an approximate 
space-time description of the scattering 
when considering localized wave packets rather than plane waves: 
for sufficiently high frequencies 
(i.e. short wave lengths 
compared to the $($acceleration of the mirror$)^{-1}$), 
the coefficients $\bt_{\om k}$ vanish
and the mean frequency of the reflected packet $\bar \om$ 
is related to $\bar k$, that of the incident one, 
by the Doppler effect
$\bar \om = \bar k \: \partial_V U_{cl}\vert_{U=\bar U}$ 
evaluated at $\bar U$, 
the retarded time of the center of the incident packet.
These two properties are easily obtained 
by evaluating the integrals 
in eq. (\ref{overlap}) by the saddle point method.

When $ \al_{\om k}$  and $ \bt_{\om k} $ are known,
the quantum scattering problem is also solved.
This follows from the linearity of  \reff{1} and \reff{2}:
when working in a second quantized framework,
being linear, these equations
provide the Heisenberg equations for the field operator.
Thus the field operator can 
be written both in the $in$ and the $out$ basis 
by
\ba \label{phi}
\Phi
&=& \intok 
\Big( 
a_k^{in}  \varphi^{in}_{k} 
+b^{in \: \dagger}_k  \varphi^{in\: *}_{k}
\Big)
\nn
&=& \intoom 
\Big( a_\om^{out}  \varphi_{\om}^{out} 
+ b^{out \: \dagger}_\om  \varphi_{\om}^{out\: *}   
\Big) \ .
\ea
When imposing that it satisfies the equal time commutation relation
$[ \Phi(z), \partial_t \Phi^\dagger(z')] = i\delta(z-z')$, 
\reff{normalization} guarantees 
that the $in$-operators $a_k , b_k$ 
satisfy the usual commutation relations 
leading to the particle interpretation.
Then, 
the $in$ vacuum $\ket{0_{in}}$ is defined 
as the product of the ground states of the $in$ oscillators 
and its excitations are generated 
by the creation operators $a^{in \: \dagger}_k, b^{in \: \dagger}_k$. 
Moreover, by construction of the $in$ modes on  $\scrym$, 
the $in$ particles 
correspond to the usual Minkowski particles on $\scrym$.
Similarly, by construction of the $out$ modes, 
all these properties apply to  
the $out$ operators $a^{out \: \dagger}_k, b^{out \: \dagger}_k$
and to the $out$ vacuum $\ket{0_{out}}$ 
when replacing $\scrym$ by $\scryp$.

Given the orthonormal and complete character 
of the $in$ and $out$ mode basis,
eqs. (\ref{overlap}) and (\ref{phi}) 
determine the Bogoliubov relations :
\ba \label{bogop1}
\left\{
\begin{array}{lr}
\disp a^{in}_k 
= \intoom ( 
\al_{\om k} \: a^{out}_\om + \bt_{\om k} \: b^{out \: \dagger}_\om 
) 
\\
\disp b^{in \: \dagger}_k 
= \intoom ( 
\bt_{\om k}^* \: a^{out}_\om + \al_{\om k}^* \: b^{out \: \dagger}_\om
)
\end{array} 
\right. 
\; \; \; \; \; 
\left\{
\begin{array}{lr} 
\disp  a^{out}_\om
= \intok ( 
\al^*_{\om k} \: a^{in}_k   - \bt_{\om k} \: b^{in \: \dagger}_k 
) 
\\
\disp b^{out \: \dagger}_\om
= \intok ( 
- \bt_{\om k}^* \: a^{in}_k  + \al_{\om k} \: b^{in \: \dagger}_k 
) \, . \ 
\end{array} 
\right. 
\ea
Then eqs. (\ref{bogcoef}) guarantee the compatibility 
of the particle interpretation in each basis, i.e., both 
 $in$ and $out$ operators obey the canonical commutations relations. 
With the relations (\ref{bogop1}), 
all questions concerning quantum scattering processes
can be answered. For instance, 
the probability amplitude to obtain a given final state 
$\ket{\Psi_{fin}}$ specified on $\scryp$ in terms of $out$ operators 
starting from some $in$ state $\ket{\Xi_{in}}$ constructed on $\scrym$
is given by the (Fock space) product 
$\scal{ \Psi_{fin}}{\Xi_{in}}$.
More intrinsic is the overlap $Z^{-1}=\scal{0_{out}}{0_{in}}$
between the initial and final vacuum states. Indeed, 
it determines the probability amplitude for the (spontaneous)
decay of the vacuum.
The computation of $Z$ is easy when the scattering is
stationary, see {\it e.g.}  \cite{BD,Primer}. 
In the general case however, due to the 
frequency mixing between $in$ and $out$ modes, 
the calculation of $Z$ is less trivial.
This generalization is presented in Appendix A.

It should also be noted that the Bogoliubov coefficients themselves are 
given by the following matrix elements
\ba
\al_{\om k}^* &=& \expect{0_{in}}{a^{out}_\om \: a^{in \: \dagger}_k}{0_{in}} \ ,
\nn
- \bt_{\om k}^* &=& \expect{0_{in}}{b^{out \: \dagger}_\om 
\: a^{in \: \dagger}_k }{0_{in}} \ .
\ea
However it is not clear how to attribute a physical meaning 
to these equations.
In particular, the relationship with the second one and pair creation
amplitude is quite obscure.
Indeed the {\it probability} amplitude to obtain on $\scryp$ one 
pair of quanta of frequencies $\om$ and $\om'$ in the $in$ vacuum 
is given by 
\be \label{apaircreat}
\expect{0_{out}}{a_\om^{out} b_{\omp}^{out}}{0_{in}}
= - { 1 \over Z} \int_0^\infty dk \: \bt_{\om k} \al^{-1}_{k \omp} \ .
\ee
We shall return to these questions of interpretation
in subsection 2.3.

Instead of considering $in-out$ matrix elements in Fock space, 
more attention has been put on expectation values of (local) 
operators in a given initial state.
The most studied object is probably the energy flux 
emitted by the mirror 
when the state of the field is $in$ vacuum. 
The motivations for this analysis are, 
first, its relevance for black hole radiation\cite{BD}-\cite{P:hawkfeynman}, 
secondly, that its non-vanishing value is due to spontaneous 
pair creation, a specific feature of quantum field theory (QFT),
and thirdly that this value can be computed both from
using eqs. (\ref{bogop1})
and from the properties of the Green function of the field.
  
Having at our disposal the Bogoliubov coefficients 
$\al_{\om k}, \bt_{\om k}$,
we start with the particle point of view.
We consider the density energy of the emitted flux. The 
corresponding hermitian operator is\footnote{
The symmetrization is due to the fact that we deal with a complex field.
Of course, in the DF model, 
particles and anti-particles equally contribute 
to $\ave{T_{VV}}$. 
This explains the overall factors of 2 in the next equations.
We warn the reader that this equal contribution will not be necessarily
found when considering partially transmitting mirrors.}
$T_{VV}=\div \Phi^\dagger \div \Phi + \div \Phi \div \Phi^\dagger$.
On the left of the mirror ($U > U_{cl}(V)$), 
using \reff{bogop1} and the first line of \reff{bogcoef}, 
the expectation value of the energy flux 
is
\ba \label{TVV2}
\ave{T_{VV}}
&\equiv& 
\expect{0_{in}}{\TVV}{0_{in}} - \expect{0_{out}}{\TVV}{0_{out}}
\nn
&=& 
2 \: \re \left[  \int\!\!\!\!\intoom \! d\omp 
{\sqrt{\om \om^\p} \over 2 \pi} \right.
\nn && 
\quad \quad
\left.
\Big[ \: 
e^{-i( \omp -\om)V }
\left( 
\intok
\bt_{\om k}^* \bt_{\omp k} 
\right)
-  
e^{-i (\omp + \om)V} 
\left( 
\intok 
\al^*_{\om k} \bt_{\omp k} 
\right) 
\Big] \right] \, .\quad \quad
\ea
It should be noted that the subtraction 
of the $out$ vacuum flux 
follows from the prescription of subtracting 
the contribution of the Minkowski vacuum. Indeed,
by construction of the $out$ modes, 
they coincide with the usual Minkowski modes on $\scryp$.

The total energy emitted to $\scryp$ is 
obtained from integrating $\ave{T_{VV}}$ over $V$. 
The integration eliminates 
the second term which is due to interferences 
between states with different particle numbers. It gives
\ba \label{H}
\ave{H_V} 
&=& 
\intev \ave{T_{VV}}
\nn
&=& 
2  \disp \intoom \om \intok {|\bt_{\om k}|}^2
= 2 \intoom \om \, \ave{n_{\om}} \ .
\ea
One gets the usual relationship between the mean energy 
and the mean number of particles 
$\ave{n_{\om}}= \intok {|\bt_{\om k}|}^2$ found on $\scryp$
(it equals to the number of anti-particles).
In this writing one sees 
that the non-vanishing character of $\ave{H_V}$
is due to the $\bt$ coefficients 
which govern the vacuum decay, see eq. (\ref{vacdecay})
in App. A.  

We now reconsider the flux $\ave{T_{VV}}$ 
without making use of the Bogoliubov coefficients 
and with less emphasis on the notion of particle.
This method is based on the two-point Wightman function
evaluated in the $in$ vacuum 
\ba
\expect{0_{in}}{\Phi^\dagger(U,V) \: \Phi(U',V')}{0_{in}} 
&=& 
\intok \varphi_k^{in}(U,V) \varphi_k^{in\, *}(U',V') \ .
\ea
In terms of this function, using eq. (\ref{scatt}),
the mean flux on $\scryp$ reads
\ba \label{TVVDFlocal}
\ave {T_{VV}} 
&=&  
2 \lim_{V' \to V} 
\left[ 
\expect{0_{in}} { \partial_{V} \Phi^\dagger \partial_{V'} \Phi}{0_{in}}
- \expect{0_{out}}{ \partial_{V} \Phi^\dagger \partial_{V'}\Phi}{0_{out}}
\right]
\nn
&=&  
-{1 \over 2\pi}\lim_{V' \to V}
\partial_{V} \partial_{V'} 
\left[ 
\mbox{ln} | U_{cl}(V') - U_{cl}(V) | 
- \mbox{ln} | V' - V | 
\right]
\nn 
&=&
\frac{1}{6\pi}
\left[
\left(
\frac{dU_{cl}}{dV}
\right)^{1/2}
\di_V^2
\left(
\left(
\frac{dU_{cl}}{dV}
\right)^{-1/2}
\right)
\right]
\nn
&=& 
{1 \over 24 \pi} 
\left[ 
\left(
{ d^2U_{cl}\over dV^2} 
\right) 
\left( 
{dU_{cl} \over dV }
\right)^{-1} 
\right]^2
-{1 \over 12 \pi} \: \partial_V 
\left[ 
\left(
{d^2U_{cl}\over dV^2 }
\right) 
\left( 
{dU_{cl} \over dV} 
\right)^{-1} 
\right]\ .
\ea
Again, the subtraction of the $out$ vacuum flux follows 
from the prescription of subtracting the 
contribution of the Minkowski vacuum.
In this second description of the flux, 
it is through this prescription 
that the notion of vacuum decay is re-introduced. 
Indeed, on $\scryp$,
the above subtraction is equivalent to normal ordering with 
respect to $out$ operators.
(This is straightforwardly proven by using eq. (\ref{bogcoef}).)
Moreover, the fact that $\ave {T_{VV}}$
identically vanishes for inertial trajectories, 
i.e. when $\partial^2_V U_{cl} =0$, 
confirms that its non-vanishing character
is due to the non-adiabaticity\cite{wdwpc} of the scattering,
a notion deeply rooted to the spontaneous creation of pair of
particles.

>From \reff{TVVDFlocal} we learn that the energy flux is local 
in that it depends only on three derivatives of the trajectory $U_{cl}(V)$ 
evaluated at the retarded time $V$
(remember that we are on the left of the mirror). 
We shall see below that this locality is a consequence of 
dealing with a perfectly reflecting mirror 
for all frequencies.

Notice finally that in \reff{TVV2}, 
the first term is positive definite 
and leads to the positive total energy $\ave{H_V}$. 
Indeed, being a total derivative, 
the second term does not contribute to $\ave{H_V}$ 
when the trajectory is asymptotically inertial 
since $U_{cl}(V) \sim V$ for asymptotically late and early $V$'s.
This might not be the case for trajectories 
which enter or leave the space through the null infinities, 
because of the infinite Doppler effect encountered asymptotically.

\subsection{Partially transmitting mirrors}

\par
In this subsection 
we study partially transmitting (but still recoil-less) non-inertial mirrors.
We shall proceed in three steps. 
We first show that the scattering by a non-inertial mirror 
is most simply described in terms of the wave functions 
which are eigen-modes of the proper time of the mirror.
(We shall call them the proper-time modes.)
When using these modes, 
the matrix relating the scattered modes to the initial ones 
is diagonal in the frequency,
exactly like for the scattering of Minkowski modes by a mirror at rest. 
Secondly, we shall see that these modes are well adapted 
to introduce partially reflecting coefficients 
with arbitrary frequency dependent phase and amplitude. 
Indeed, since this matrix is diagonal in the proper-time frequency, 
unitarity constraints its elements 
in a simple manner, frequency by frequency. 
Thirdly, for both partial and total reflection, 
we shall see that
the usual Bogoliubov coefficients, eq. (\ref{overlap}), 
relating the $in$ and $out$ Minkowski modes 
are simply obtained from this diagonal matrix. 


To fulfill this program, 
we first need to construct the proper-time modes. 
To this end it is very useful to introduce new light-like coordinates $u, v$
such that the time-like coordinate $(u+v)/2 = \tau$ 
is the proper time of the mirror 
and the space-like one defined by $(v-u)/2 = \rho$ 
is such that the trajectory reads $\rho = \rho_0= constant$.
These new coordinates are defined by two analytic functions 
$u(U)$ and $v(V)$ where $U, V$ are the Minkowski light-like coordinates.
These functions are determined by the mirror trajectory $U_{cl}(V)$
and the two conditions defining $\tau$ and $\rho$. 
Indeed, along the mirror's trajectory, 
the length element obeys
\ba
ds^2 
&=& 
\partial_V U_{cl}(V) \; dV^2 
= \partial_U V_{cl}(U) \; dU^2
\nn
&=& 
dv^2 =  du^2 = d\tau^2 \ .
\ea 
This gives
\ba
\frac{dv}{dV} 
= \sqrt{\partial_V U_{cl}} \quad
{\rm and} \quad 
\frac{du}{dU} 
= \sqrt{\partial_U V_{cl}}
\ .
\ea
One verifies that the Jacobians 
${dv}/{dV}$ and ${du}/{dU}$ 
define a time dependent boost 
since they satisfy 
$({dv}/{dV})({du}/{dU}) =1$ for all $\tau$.
The proper-time modes are then simply given by
\ba
\label{prtm}
\varphi_\la(u) 
&=& 
\frac{e^{-i \la u}}{\sqrt{4 \pi |\la|}} \ ,
\nn
\varphi_\la(v) 
&=&
\disp \frac{e^{-i \la v}}{\sqrt{4 \pi |\la|}} \ .
\ea
They form a complete basis on 
$\scrym$ and $\scryp$ 
since our condition 
that the trajectory emerges from $i^-$ and finishes
on $i^+$ 
implies that the $v$ and $u$ axes 
cover those of $V$ and $U$ respectively.

In the case of total reflection,
the scattering along the mirror at $\rho = \rho_0$ is trivial.
When using the conventions of the former subsection 
(eqs. (\ref{inmodes})  and (\ref{outmodes})), 
one has, on the left of the mirror,
\ba
\varphi_\la^{U, in}(u,v) 
&=& 
\varphi_\la(u) - e^{ 2i \la \rho_0} \, \varphi_\la(v) 
\nn
&=& 
- e^{ 2i \la \rho_0} \, \varphi_\la^{V, out}(u,v) \ .
\ea
The new subscript $U, V$ indicates which side of $\scrym$ ($\scryp$) 
is the  asymptotic  support of the $in$ ($out$) functions. 
We have introduced it 
in order to describe partial reflection 
which requires to consider simultaneously both sides
of the mirror.
Using this notation, on the right of the mirror, one has
\ba
\varphi_\la^{V, in}(u,v) 
&=& 
\varphi_\la(v) - e^{ - 2i \la \rho_0} \, \varphi_\la(u) 
\nn
&=& 
- e^{ -2i \la \rho_0} \, \varphi_\la^{U, out}(u,v) \ .
\ea
It will be useful to express these relations 
by a $2\times 2$ matrix $\S_\la$ as
\be \label{defS}
\varphi_\la^{i, out} 
= S_\la^{ij} \varphi_\la^{j, in}
\quad 
\Big( 
\equiv a^{j, in}_\la 
= S_\la^{ij} a^{i, out}_\la
\Big)\ .
\ee
At fixed $\la$, 
the indices of rows and columns $i, j$ 
are the new subscript $U$ or $V$. As usual,
 repeated indices are summed over.
For total reflection, one has
\ba \label{Srefltot}
\S_\la 
= 
\left(
\begin{array}{cc} 
0 & -\: e^{+2i \la \rho_0} \\
-\: e^{-2i \la \rho_0} & 0 
\end{array}
\right) \ .
\ea

We now consider partial reflection.
When considering elastic reflection, 
the matrix $\S_\la$ relating $in$ and $out$ modes 
which generalizes \reff{defS} is unitary.
(That is, we generalize total reflection in a restricted way since
we keep both the linearity and the unitarity of \reff{defS}.)
Unitarity constraints the elements of $\S_\la$ 
\ba
\S_\la
= \; 
\left(
\begin{array}{cc} 
s_u \: e^{i\varphi_u} & -i \: R \:  e^{i\varphi} \\
-i\: R' \:  e^{i\varphi'} & s_v \: e^{i\varphi_v} 
\end{array}
\right) 
\ea
to obey
\ba
R=R' &,& \quad s_u = s_v \, ,
\nn 
s_u^2 + R^2 = 1 &\and& 
\varphi' = \varphi_u + \varphi_v - \varphi \, .
\ea
(For simplicity of the expressions, 
we have not written the argument $\la$ 
but all variables should be understood as $\la$ dependent.)
Physically, $R$ and $s$ correspond to the reflection and 
transmission coefficients, 
\ie when working in the rest frame of the mirror,  
the {probability} for an incident quantum of frequency $\la$
to be reflected is $R^2$. 

In what follows we impose $\varphi_u = \varphi_v= -\phi$, 
a condition which expresses 
that the transmitted part of scattering 
is independent of the sign of the momentum. 
In anticipation to Section $3$, we point out 
that this equality is automatically satisfied 
when considering parity invariant hamiltonians 
(see \cite{IZ} Chap. 3.4). 
In this case the matrix reads
\ba \label{Stype}
\S_\la = 
e^{-i\phi} 
\left(
\begin{array}{cc} 
\sqrt{1-R^2} & -i \: {R}
\:  e^{i\theta} \\
-i \: 
{R}
\:  e^{-i\theta} & \sqrt{1-R^2}
\end{array}
\right) \ .
\ea
In principle the common phase $e^{-i\phi}$ could be 
re-absorbed in a redefinition of the modes. 
However, when using $in$ and $out$ modes conventionally defined,
\ie $\varphi_\la^{V, in}(v) =  \varphi_\la^{V, out}(v)
=\varphi_\la(v)$ of \reff{prtm}, 
the phase $\phi$ is univocally fixed.
As we shall see in the next Section,
this convention is automatically used when 
considering interactions perturbatively.
This is also the case in the DF model. Indeed
the limiting case of total reflection given in
 \reff{Srefltot}
is reached for $R \to 1$ and $\phi = \pi/2$ for all $\la$.
One also finds that the other phase $\theta$ 
is related to the mirror location by $\theta= 2 \rho_0 \la$. 

To complete our second step, 
we should describe particles and anti-particles
simultaneously. 
To this end, we group the $in$ operators 
$\big(a_\la^{U, in}, a_\la^{V, in}, 
b_\la^{U, in \: \dagger}, 
b_\la^{V, in \: \dagger}\big)$
in a $4$-vector $a_\la^{\mu, in}$
and the $out$ operators 
$\big(a_\la^{U, out}, a_\la^{V, out}, 
b_\la^{U, out \: \dagger} ,
b_\la^{V, out \: \dagger}\big)$
in  $a_\la^{\mu, out}$. 
Similarly, we group their corresponding modes in
the $4$-vectors $\varphi_\la^{\mu, in}$ and $\varphi_\la^{\mu, out}$.
Since we work with a charged field, 
the modes associated with 
$b_\la^{i \: \dagger}$ might not be the complex conjugates
of those associated with $a_\la^i$. (As it is the case 
when dealing with a charged field in an electro-magnetic field,
see {\it e.g.} Sect. 1.3 in \cite{Primer}).
Explicitly, in our case, the $4$ modes are 
$\big(\varphi_\la^{U}, \varphi_\la^{V}, 
\bar{\varphi}_\la^{U \: *},
\bar{\varphi}_\la^{V \: *}\big)$
where $\bar{\varphi}^i_\la$ 
designate the two modes associated with anti-particles 
operators $b^U_\la$ and $b^V_\la$.  

We then introduce the $4\times 4$ matrix 
given by
\ba \label{Slala'}
\S_{\la \la'} 
= \delta( \la -\la')
\left(
\begin{array}{cc} 
\S_\la &  0\\
0 & \bar{\S}^*_{\la}
\end{array}
\right) \, .
\ea
where $\bar{\S}_{\la}$ is the scattering matrix 
for the anti-particles. 
Since $\S_{\la\lap}$ is block diagonal,
unitarity constrains ${\S}_{\la}$ and $\bar{\S}_{\la}$
separately. 
$\S_{\la\lap}$ acts on the $in$ $4$-vector as follows
\be \label{newrel}
\varphi_\la^{\mu, out } 
= S_{\la \lap}^{\mu \nu} \, \varphi_{\lap}^{\nu, in } 
\ee
where continuous repeated indices are integrated from $0 \to \infty$
and discrete ones summed over the four components 
defined at fixed frequency. 
With these choices, 
the components of $\S_{\la\la'}$ are the Bogoliubov coefficients 
conventionally defined. By conventionally defined we mean
the equations which generalize \reff{bogop1}, i.e., 
\ba
\label{newbogrel}
a^{j, in}_{\lap} &=& \al^{ij}_{\la \lap} \: a^{i , out}_{\la} 
+ \btb^{ij}_{\la \lap} \: b^{i , out \: \dagger}_{\la} ,
 \nn
b^{j, in\, \dagger}_{\lap}  &=& \bt^{ij \: *}_{\la \lap} \: a^{i , out}_{\la}
+ \alb^{ij \: *}_{\la \lap} \: b^{i , out \: \dagger}_{\la} \ ,
\ea
where the Bogoliubov coefficients 
$\al,\alb,\bt,\btb$
are now $2 \times 2$ matrices. 
By direct identification, one obtains
\ba \label{defbogoij}
\al^{ij}_{\la \lap} 
&=& \scal{\varphi_{\lap}^{j, in}}{\varphi_{\la}^{i, out}}
= S^{ij}_{\la \lap}\ , 
\nn
{\alb}^{ij}_{\la \lap} 
&=& \scal{\bar{\varphi}_{\lap}^{j, in}}{\bar{\varphi}_{\la}^{i, out}}
= {S^{i+2 \, j+2}_{\la \lap}}^{\: *}\ ,
\nn
\bt^{ij}_{\la  \lap} 
&=& \scal{\bar\varphi_{\lap}^{j, in}}{\varphi_{\la}^{i, out \: *}}
= {S^{i\,j+2}_{\la \lap}}^{\: *}\ , 
\nn
{\btb}^{ij}_{\la  \lap} 
&=&\scal{\varphi_{\lap}^{j, in}}{\bar{\varphi}_{\la}^{i, out \: *}}
= S^{i+2 \,j}_{\la \lap} \ .
\ea 
When $\S_{\la \la'}$ is block-diagonal
in the sense of \reff{Slala'}, 
one obviously has $\bt^{ij}_{\la  \lap}={\btb}^{ij}_{\la  \lap}=0$.
In full generality, 
$\S_{\la \la'}$ satisfies unitarity in the following sense
\be
(S^\dagger)_{\la \la''}^{\mu \nu} \, S_{\la'' \la'}^{\nu  \mu'} 
= \delta(\la - \lap) \, \delta^{\mu \mu'}\ .
\ee
This equation generalizes eqs. (\ref{bogcoef}) to partially
transmitting mirrors.

With eqs. (\ref{Srefltot}), (\ref{Slala'}) and (\ref{newrel}), 
we have shown that the scattering in the DF model
is trivial when using the proper-time modes. 
We have done more since 
eqs. (\ref{newrel}) and (\ref{defbogoij}) apply 
to all partially transmitting mirrors governed by $\S_\la$ given by 
\reff{Stype}.

The last step consists in finding the relationship 
between $\S_{\la \la'}$ and the Bogoliubov coefficients 
between $in$ and $out$ Minkowski modes.
This is simply achieved by introducing the $4 \times 4$ matrix 
which relates the (unscattered) Minkowski modes of frequency $k = -i  \di_t$ 
to the (unscattered) proper time modes of frequency $\la = -i  \di_\t$ :
\be
\phi_k^{\mu} 
= {\cal{B}}^{\mu \nu }_{k \la} \varphi_\la^{\nu}\ .
\ee  
The elements of this matrix are given by
\ba \label{bigB}
{\cal{B}}_{k \la} 
&=& 
\left(
\begin{array}{cccc} 
\scal{\varphi^{U}_{\la}}{\phi^{U}_{k}} & 0
& -\scal{{\varphi^{U \: *}_{\la}}}{\phi^{U}_{k}} &  0 \\
0 &\scal{\varphi^{V}_{\la}}{\phi^{V}_{k}}  &0
& - \scal{\varphi^{V \: *}_{\la}}{\phi^{V}_{k}}  \\
\scal{\varphi^{U}_{\la}}{\phi^{U \: *}_{k}} & 0
& - \scal{\varphi^{U \: *}_{\la}}{\phi^{U \: *}_{k}}  &0  \\
0 &\scal{\varphi^{V}_{\la}}{\phi^{V \: *}_{k}} & 0 
& - \scal{\varphi^{V \: *}_{\la}}{\phi^{V \: *}_{k}}  
\end{array}
\right) 
\nn
&=& 
\left(
\begin{array}{cccc} 
\al^{UU}_{k  \la} & 0 & \bt^{UU \: *}_{k \la} & 0 \\
0 & \al^{VV}_{k \la} & 0 & \bt^{VV \: *}_{k \la} \\
\bt^{UU}_{k \la} & 0 & \al^{UU \: *}_{k \la} & 0 \\
0 & \bt^{VV}_{k \la} & 0 & \al^{VV\: *}_{k \la}
\end{array}\right)\ .
\ea
Since ${\cal{B}}_{k \la}$ relates unscattered modes,
it is independent of the charge of the particle, 
hence ${\cal{B}}_{k \la}^{1\, 1} \equiv \al^{UU \:}_{k \la}
=\bar{\al}^{UU}_{k \la}  \equiv 
{\cal{B}}_{k \la}^{3\, 3}$. The same equality applies 
to  $\al^{VV}_{k \la}, \bt^{VV}_{k \la}, \and \bt^{UU}_{k \la}$.

The important point for us is that 
 ${\cal{B}}_{k \la} $ 
also relates the $in$ Minkowski modes to the $in$ proper-time modes
and the $out$ Minkowski modes to the $out$ proper-time modes.
Therefore the linear relation between $in$ and $out$ Minkowski modes 
is given by 
\be \label{meaneq1}
\phi_\om^{\mu, out} 
=
S^{\mu \mu'}_{\om k} \; \phi_k^{\mu', in} 
\ee
where
\ba
S^{\mu \mu'}_{\om k}
&=& 
{\cal{B}}^{\mu \nu}_{\om \la} \, 
S^{\nu \nu'}_{\la \lap} \, 
({{\cal{B}}_{k \lap}}^{-1})^{\nu' \mu'} \, .
\ea
Repeated indices are summed over and 
the inverse of ${\cal{B}}$ is defined by
\be
\varphi_{\la}^{\nu} 
= ({{\cal{B}}_{k \la}}^{-1})^{\nu \mu} \phi_{k}^{\mu}
\, .
\ee  
It is given by
\ba
{\cal{B}}_{k \la}^{-1} = 
\left(
\begin{array}{cccc} 
\al^{UU \: *}_{k \la} & 0 & -\bt^{UU \: *}_{k \la} & 0 \\
0 & \al^{VV\: *}_{k \la} & 0 & -\bt^{VV \: *}_{k \la} \\
-\bt^{UU}_{k \la} & 0 & \al^{UU}_{k \la} & 0 \\
0 & -\bt^{VV}_{k \la} & 0 & \al^{VV}_{k \la} 
\end{array}
\right)\, .
\ea
Explicitly, using the dictionary (\ref{defbogoij}) now applied to
Minkowski modes, 
the four coefficients  $S^{1\nu}_{\om k}$ are
\ba \label{meaneq2}
\al^{UU}_{\om k} 
&=& 
\delta(\om - k) 
-i \intola \big(\al^{UU}_{\om \la} T_\la^{UU} \al^{UU \: *}_{k \la}
              + \bt^{UU \: *}_{\om \la} \bar{T}_\la^{UU \: *} \bt^{UU}_{k \la} \big) \ ,
\nn
\al^{UV}_{\om k} 
&=& 
-i \intola \big(\al^{UU}_{\om \la} T_\la^{UV} \al^{VV \: *}_{k \la}
              + \bt^{UU \: *}_{\om \la} \bar{T}_\la^{UV \: *} \bt^{VV}_{k \la} \big) \ ,
\nn
\bt^{UU \: *}_{\om k}
&=& i \intola \big(\al^{UU}_{\om \la} T_\la^{UU} \bt^{UU \: *}_{k \la}
                 + \bt^{UU \: *}_{\om \la} \bar{T}_\la^{UU \: *} \al^{UU}_{k \la} \big) \ ,
\nn
\bt^{UV \: *}_{\om k}
&=&
i \intola \big(\al^{UU}_{\om \la} T_\la^{UV} \bt^{VV \: *}_{k \la}
             + \bt^{UU \: *}_{\om \la} \bar{T}_\la^{UV \: *} \al^{VV}_{k \la} \big) \ .
\ea
Similar equations give the expressions for the remaining components
of $S^{\mu \nu}_{\om k}$.
We have written $\S_\la$ as $\S_\la = \1 -  i \: \T_\la$ 
(and $\bar{\S}_\la= \1 -  i \: \bar\T_\la$)
in order to extract the trivial part of the diagonal elements.
This trivial part leads to the delta function in the first equation.
The usefulness of the writing is that
it will be easily related to the perturbative expressions
we shall encounter in the next Section.

Equations (\ref{meaneq2})
are the central result of this Section.
They give the $in-out$ overlaps of Minkowski modes 
in terms of the matrices $\T_\la, \bar{\T}_\la$
computed in the rest frame of the mirror 
and the overlaps between the free (unscattered) 
Minkowski and proper time modes.

It is then easy to obtain the mean flux 
emitted by this partially transmitting non-inertial mirror 
when the initial state of the field is the Minkowski vacuum. 
The same algebra which gave \reff{TVV2} now gives
\ba
\label{newTvv0}
\ave{\TVV} = \ave{\TVV}^{particle} + \ave{\TVV}^{anti-particle} \ ,
\ea
where
\ba \label{newTvv}
\ave{\TVV}^{particle}
&=& \re  \left[ \sum_{j = U,V}
\int \!\!\!\!\intoom \! d\omp  {\sqrt{\om \omp} \over  2 \pi} 
\right.
\nn 
&& \quad \!  \left. \left[
e^{-i( \omp -\om)V }
\left( 
\intok \bt^{Vj \: *}_{\om k} \: \bt^{Vj}_{\omp k} 
\right) 
-
e^{-i (\omp + \om)V} 
\left(  
\intok \bar\al^{Vj \: *}_{\om k} \: \bt^{Vj}_{\omp k} 
\right) 
\right]\right]  .\quad \quad
\ea
$\ave{\TVV}_{anti-particle}$ is given by
the same expression 
with $\alb,\bt$ replaced by $\al,\btb$. 
$\ave{\TVV}$ possesses the same structure as \reff{TVV2}. 
However, four kinds of coefficients $\al, \bt$ 
should be considered since we are 
dealing with {\em partial} reflection of {\em charged} particles.

When the scattering is independent of the energy and the charge 
of the particles \ie when
$R$ and $\phi$ defined in \reff{Stype}
are independent of $\la$
and when $\bar{\S}_\la^* = \S_{-\la}$,
the integration over $\la$ can be trivially performed 
as it expresses the completeness of the $\varphi_\la$ modes. 
In this case, 
as in the DF model, one has $\bt^{UU}_{\om k}=\bt^{VV}_{\om k}= 0$. 
One also finds that the emitted flux is simply 
\be
\ave{\TVV} = R^2 \: \ave{\TVV}_{DF} \ ,
\ee
where $\ave{\TVV}_{DF}$ 
is the flux found in the DF model, see (\ref{TVV2}).

Instead, when $R$ and $\phi$ depend on 
the energy and/of the charge, 
$\bt^{UU}_{\om k}$ and $\bt^{VV}_{\om k}$
will be, in general\footnote{$\bt^{UU}_{\om k} = 0$
requires that $\bar T^{UU \: *}_{\la} = - T^{UU}_{-\la}$
for all $\la >0$, 
and similarly for the $VV$ coefficients. 
In the next Section, we shall see that condition is satisfied 
for the time independent couplings with U(1) symmetry.}, 
different from zero.
In this case, one also looses the possibility of re-expressing the flux
in terms of the derivatives of the trajectory as we did
it in \reff{TVVDFlocal}.
This can be understood from eqs. (\ref{meaneq2}) :
when expressing 
$\T_\la$ as a series in powers of $\la$, 
one would obtain for $\ave{\TVV}$ 
a non-local expression in $V$
unless the series in $\la$ stops after a finite number 
of terms.

\subsection{Additional remarks}

In this subsection, we relate the matrices
$\S_{\la \la'}$ and $\S_{\om k}$ which act
linearly on $in$ and $out$ operators
to the conventional $S$ matrix acting on multi-particle 
states in the Fock space.
With this identification we shall be able to relate the 
Bogoliubov coefficients \reff{meaneq2} 
to {\it transition amplitudes} and not only to
expectation values as in \reff{newTvv}. 

By definition\cite{IZ}, the action of this operator 
on states and operators is the following 
\ba
\label{Sso}
\ket{0_{in}} \!
&=& \!  \Ss \: \ket{0_{out}}, 
\nn
a^{i, out}_{\la}  =  \Ss^{-1}  \; a^{i, in}_{\la} \;  \Ss  &,&  
b^{i, out \; \dagger}_{\la}  =  
\Ss^{-1}   \; b^{i, in \; \dagger}_{\la} \;  \Ss \ .
\ea
Since we are dealing with elastic scattering,
this operator contains exactly the same information 
as the matrices $\S_\la, \bar \S_\la$.
Indeed,
the block diagonal character of \reff{Slala'} 
and the linearity of \reff{newrel} tell us that $\Ss$ 
is the exponential of a quadratic form of 
proper-time operators $a_\la, b_\la$: 
\ba \label{Shat}
\Ss 
= e^{ \disp -i  
\big( 
a^{i, in}_\la s^{ij}_{\la \lap} a^{j, in \: \dagger}_{\lap} 
- b^{i, in \: \dagger}_\la \bar s^{ij}_{\la \lap} b^{j, in}_{\lap} 
\big) } \ .
\ea
Then straightforward algebra gives
\ba
{\mathbf s}_{\la \lap}
&=& \delta(\la-\lap) \: 
\left(
\begin{array}{cc}
\phi & \mbox{Arcsin}(R) e^{i\theta} \\
\mbox{Arcsin}(R) e^{-i\theta} & \phi 
\end{array}
\right) \ , \nn
\bar {\mathbf s}_{\la \lap}
&=&
{\mathbf s}_{\la \lap}(\bar R,\bar \theta,\bar \phi) \ ,
\ea
where $R,\theta,\phi$ have been defined 
in \reff{Stype} and $\bar R,\bar \theta,\bar \phi$ are defined 
in the same way from $\bar \S_\la$.
We note that in the DF model, i.e. in the limit of perfect reflection,
${\mathbf s}_{\la \lap}$ is given by
\be
  {\mathbf s}^{DF}_{\la \lap}
= \delta(\la-\lap)\, ({\pi \over 2}) \: 
\left(
\begin{array}{cc}
1 &  e^{i\theta} \\
 e^{-i\theta} & 1
\end{array}
\right) = \bar {\mathbf s}_{\la \lap}^{DF}\, . 
\ee
Although the configurations on the left and and on the
right of the mirror completely decouple, the S-matrix $\Ss$ 
treats both sides simultaneously. 

To anticipate the expression of $\Ss$ in terms of Minkowski operators
which will mix creation and destruction operators,
it is a convenient to rewrite \reff{Shat} in term of the  
$4$-vector $a_\la^{\mu, in}$:
\ba \label{Shatshort}
\Ss = e^{ \disp - i  
\big( 
a^{\mu, in}_\la s^{\mu\nu}_{\la \lap} a^{\nu, in \: \dagger}_{\lap}  
\big) } 
\; \; \with \; 
(s^{\mu\nu}_{\la \lap}) 
= \left(
\begin{array}{cc}
s^{ij}_{\la \la'} & 0 \\
0 &  - \bar{s}^{ij}_{\la \la'}
\end{array}
\right) \ .
\ea

To obtain the expression of $\Ss$ in terms of 
the Minkowski operators $a^{i, in}_k, b^{i, in}_k$, 
it suffices to use the matrix ${\cal{B}}_{\om \la}$
to replace proper-time operators by Minkowski ones.
Explicitly one obtains
\ba
\Ss =  e^{ \disp - i  
\big( 
a^{\mu, in}_\om s^{\mu\nu}_{\om \omp} a^{\nu, in \: \dagger}_{\omp}
\big) } 
\; \; \with \; 
s^{\mu\nu}_{\om \omp} 
= {\cal{B}}_{\om \la}^{\mu \mu'} \,
  s^{\mu' \nu'}_{\la \lap} \, 
  ({\cal{B}^\dagger}_{\omp \lap})^{\nu \nu'}\ .
\ea

Formally, $\Ss$ provides the answer 
to all questions concerning asymptotic states 
and expectation values.
For instance, the probability amplitude governing the (Minkowski) 
vacuum decay, \reff{vacdecay}, is simply
\be
Z^{-1} = \scal{0_{out}}{0_{in}} = \expect{0_{in}}{\Ss}{0_{in}} \ . 
\ee
Similarly, the probability amplitude
for an initial quantum of momentum $k$ to be scattered
and for no pair to be created is  
\be
\label{physa}
\expect{0_{out}}{a_\om^{i,out} a_{k}^{j,in \, \dagger}}{0_{in}}
= \expect{0_{in}}{a_\om^{i,in} \: \Ss \: a_{k}^{j,in \, \dagger}}{0_{in}}
= {1 \over Z } {(\al^{-1})}^{ji}_{k \om}\, .
\ee
The last equality is easily obtained by using \reff{newbogrel}
to express $a_\om^{i,out}$ in terms of $a_k^{j, in}$ and 
$b_\om^{j,out \, \dagger}$.
In the same way, the Bogoliubov coefficient $\beta$ 
is related to probability amplitude 
to find a pair of $out$ quanta in the $in$ vacuum by 
\be
\label{physb}
\bt^{ij}_{\om k} \: {(\alb^{-1})}^{ji'}_{k \omp} =
- { \expect{0_{out}}{a_\om^{i,out} b_{\omp}^{i',out}}{0_{in}}
\over \scal{0_{out}}{0_{in}}}
= - {\expect{0_{in}}{a_\om^{i,in} b_{\omp}^{i',in} \: \Ss}{0_{in}}
\over \expect{0_{in}}{\Ss}{0_{in}}} \ .
\ee
It should be stressed that
these relations {\it determine} the
physical interpretation of the overlaps $\al, \beta$ given
in \reff{meaneq2}.
In fact the second quantized framework was never
used to obtain \reff{meaneq2}: 
only the linearity of the relations and the 
orthonormal character of the proper time and the Minkowski
modes basis were exploited. 

The physical interpretation of $\al, \beta$ is the following:
to first order in the transfer matrix $\T_\la$,
$\al$ ($\beta$) divided by $Z$ gives the probability amplitude
to scatter a quantum (to produce a pair of quanta), since 
$\al^{-1}\simeq 1+ iT$ ($\bt \simeq -iT$).
Upon considering higher order terms in $\T_\la$, one looses
the simplicity of the relationship so as to get the above 
equations.
The simple relation in the linear regime 
will be nicely confirmed in the next Section, when using 
perturbation theory.
We shall see in particular that the division by $Z$ 
corresponds to the usual restriction of keeping only 
the connected graphs engendered by the development of 
$\Ss = T e^{-ig\int dt H}$ in powers of $g$. 
We shall further comment on these aspects at the end of Section 3.



\section{The self-interacting model}

\par
In this section we introduce a
model based on self-interactions  
which derives from an action principle\cite{mt,Jaekel2,recmir}. 
In a first part we consider time independent couplings. 
In this case, re-summing the Born series
leads to diagonal matrices in the proper-time energy $\la$  
with parameters $R$ and $\phi$ 
which depend on $\la$ according to the number of derivatives
in the interaction hamiltonian.
This model will be generalized by considering
a thick mirror with a non-zero width.
Using a perturbative approach, 
we shall see that the thickness acts as a UV cutoff.

In the second part, we work with time dependent couplings. 
We shall work perturbatively, up to the second order in the interactions.
The novelty concerns the transients induced by the 
switching on and off of the coupling. 

\subsection{The scattering with $g$ constant}
\par
To exploit the fact that the coupling is $\tau$ independent, 
it is convenient 
to work with the coordinates $(\tau, \rho)$ 
in which the mirror is at rest.
In these coordinates, the interaction Lagrangian reads :
\ba \label{lagrangian}
L_{int} = g  
\int \!\!\! \intetau \! d\rho \;  f(\rho) \; 
J\!\left(\Phi(\tau,\rho),\Phi^\dagger(\tau,\rho)\right).
\ea
$g$ is the coupling parameter, 
$f$ is a real function which specifies the thickness of the mirror
and which is normalized as follows $\interho f(\rho) = 1$. 
$J$ is an hermitian operator which is quadratic in the complex field.
We shall consider three different cases: 
$ \Phi^\dagger \Phi + \Phi \Phi^\dagger$, 
$\Phi^\dagger i \raise 0.05mm 
\hbox{$\stackrel{\leftrightarrow}{\di_\tau}$} \Phi$,
and $\di_\tau \Phi^\dagger \di_\tau \Phi +  \di_\tau \Phi \di_\tau \Phi^\dagger$.
In the following equations, we shall present the details
only with the second expression.
At the end of the derivation,
we shall give the final results for the two other cases.

Given \reff{lagrangian}, \reff{1} is now replaced by 
\ba \label{new1}
(\partial_\tau^2 -\partial_\rho^2 )  \Phi(\tau,\rho) 
= g \: f(\rho) \: 2i \: \di_\tau \Phi.
\ea
Being linear, the solution can be expressed as 
\ba \label{eqlin}
\Phi(\tau,\rho) 
\! \! &=&  \! \!
\Phi^{in}(\tau,\rho) 
+ g \, \int \!\!\!\!\intetaup \! d\rho' \;
G^{ret}(\tau , \rho \: ; \tau', \rho') \:  
 f(\rho')  \: 2i \: \partial_{\tau'} \Phi(\tau', \rho') 
\nn
\! \! &=&  \! \!
\Phi^{out}(\tau,\rho) 
+ g \,  \int \!\!\!\!\intetaup \! d\rho' \; 
G^{adv}(\tau , \rho \: ; \tau', \rho') \:  
  f(\rho') \:  2i \: \partial_{\tau'} \Phi(\tau', \rho') \  ,
\ea
in terms of the homogeneous solution $\Phi^{in}$ ( $\Phi^{out}$)
which determines the initial (final) data.  
The retarded and advanced Green functions 
are defined, as usual, by
\ba\label{retadv}
G^{ret}(\tau , \rho \: ; \tau', \rho') 
= \int \!\!\!\!\intela dl \: \frac{1}{4\pi^2} 
\: \frac{e^{-i\la(\tau-\tau') + il(\rho-\rho')}}{l^2 - (\la+i\e)^2}
\; \; 
(=0 \; \mbox{for} \; \tau' > \tau) \ ,
\nn
G^{adv}(\tau , \rho \: ; \tau', \rho') 
=  \int \!\!\!\!\intela dl \: \frac{1}{4\pi^2} 
\: \frac{e^{-i\la(\tau-\tau') + il(\rho-\rho')}}{l^2 - (\la-i\e)^2}
\; \; 
(=0 \; \mbox{for} \; \tau' < \tau)\ .
\ea
To exploit the time independence of the coupling $g$,
we work at fixed energy with 
\be \label{phiom}
\varphi_\la(\rho) 
= \intetau {1 \over 2\pi} \:  \Phi(\tau,\rho) \: e^{i\la \tau}\ .
\ee
In Fourier transform, eqs. (\ref{eqlin}) give
\ba \label{phi+in+out} 
\varphi_\la(\rho) 
&=& 
\varphi_\la^{in}(\rho) 
+ i g \: \inte \! d\rho' \; f(\rho') \: \varphi_\la(\rho') e^{i\la|\rho-\rho'|} 
\nn
&=& 
\varphi_\la^{out}(\rho) 
- i g \: \inte \! d\rho' \; f(\rho') \: \varphi_\la(\rho') e^{-i\la|\rho-\rho'|}\ .
\ea
These equations have been obtained by using 
\ba 
\intel \frac{\disp e^{il(\rho-\rho')}}{l^2-(\la \pm i\e)^2}
= \frac{\pm 2i\pi}{ 2 (\la \pm i\e)} \: e^{\pm i\la|\rho-\rho'|}\ .
\ea

We now decompose the quantized modes $\varphi_\la^{in}$ 
in terms of creation and destruction operators,
\ba \label{phila+-}
\varphi^{in}_\la (\rho)
&=& \frac{1}{\sqrt{4 \pi \la}} 
(\: a^ {U, in}_\la \: e^{i\la\rho} 
+ a^{V , in}_\la \: e^{-i\la\rho} \: )
\; \;  \mbox{ for } \la > 0
\nn
&=& \frac{1}{\sqrt{4 \pi |\la|}} 
(\: b^{U , in  \: \dagger}_{|\la|} \: e^{-i|\la|\rho} 
+ b^{V , in \: \dagger}_{|\la|} \: e^{i|\la|\rho} \: ) 
\mbox{ for } \la < 0 .
\ea
We do the same for the $out$ modes. 
Then, for $f(\rho) = \delta(\rho-\rho_0)$,  
in the limit $\e \rightarrow 0$,
eqs. (\ref{phi+in+out}) give :
\ba
\left(
\begin{array}{c}
a^{U, out}_{\la} \\
a^{V, out}_{\la} 
\end{array}
\right)
=
\frac{1}{1-ig}
\left(
\begin{array}{cccc}
1 & ig e^{-2i\la\rho_0} \\
ig e^{2i\la\rho_0} & 1 
\end{array}
\right)
\:
\left(
\begin{array}{c}
a^{U, in}_{\la} \\
a^{V, in}_{\la} 
\end{array}
\right)\ .
\ea
We recover the linear structure of $\S_\la$ 
in \reff{defS}. 
Since the unitarity of $\S_\la$ provides 
$a^{i,out}_\la = S^{ij \: *}_\la \: a^{j,in}_\la$, 
when using the definitions
of eq. (\ref{Stype}), 
we obtain 
\ba
R = \frac{g}{\sqrt{1+g^2}}  \: , 
\; \; 
\phi = \mbox{Arctan}(g) 
\; \; 
\and 
\theta =   2 \la \rho_0 \ .
\ea 
In the strong coupling limit 
(\ie for $g \rightarrow + \infty$)
one obtains total reflexion 
(\ref{Srefltot})
in a $\la$ independent manner.
This is a special feature of the coupling  
$J=\Phi^\dagger i \raise 0.1mm 
\hbox{$\stackrel{\leftrightarrow}{\di_\tau}$} \Phi$
which is associated with a dimensionless $g$.

This analysis can be repeated 
with the two other operators previously defined.
The presence or the absence of derivatives in $J$ 
modifies the IR or UV behavior of $R$.
For $\Phi^\dagger \Phi + \Phi \Phi^\dagger$ one obtains \cite{mt} : 
\ba
R_\la = \frac{g/\la}{\sqrt{1+g^2/\la^2}}  \: , 
\; \; 
\phi_\la = \mbox{Arctan}(g/\la) 
\; \; 
\and 
\theta =   2 \la \rho_0 \ . 
\ea
In this case, the mirror is totally reflecting in the IR.
This leads to strong IR divergences when considering
time dependent coupling $g$.
On the contrary, when using  
$\di_\tau\Phi^\dagger \di_\tau\Phi + \di_\tau\Phi \di_\tau\Phi^\dagger $,
we get
\ba
R_\la =  \frac{g\la}{\sqrt{1+g^2\la^2}} \: , 
\; \; 
\phi_\la = \mbox{Arctan}(g\la) 
\; \; 
\and 
\theta =   2 \la \rho_0 \ .
\ea
In this case, the mirror is transparent in the IR limit.
This useful property will be exploited in Section $4$.

One notices that the transfer matrix can be expressed 
in a general way according to the number $n$ of derivatives $\di_\tau$
in the interaction term:
\ba
\label{analT}
\T_\la
= \frac{-g \la^{n-1}}{1-ig \la^{n-1} A_\e}
\left(
\begin{array}{cc}
1 &   e^{2i\la\rho_0} \\
e^{-2i\la\rho_0} & 1 
\end{array}
\right).
\ea
In this expression, we have not taken the limit
$\e \to 0$. 
The function 
$A_\e={\la}/({\la+i\e})$ determines the
analytical properties of $\T_\la$ in the complex $\la$ plane.
The specification of the pole of $A_\e$
follows from that of $G^{ret}$ in \reff{retadv}. It 
guarantees that causality will be respected\cite{mt}. 
This crucial ingredient was missing in Section 2.2
wherein the matrix $\T_\la$ can be chosen from the outset.
In that kinematic framework, the analytical properties 
should be imposed by hand if one wishes to implement 
causality. On the contrary, in the present case
causality follows from the Heisenberg equations. 


Eqs. (\ref{phi+in+out}) and (\ref{phila+-}) also determine
the relation between the anti-particle
$in$ and $out$ operators $b^{i \: \dagger}_\la$. 
By direct computation one finds 
$\bar\T_\la^*= - \T_{-\la}$. 
This is precisely the condition which 
gives $\beta^{UU}_{\om k}=\beta^{VV}_{\om k}=0$,
see the former footnote. 
When using $\T_\la, \bar \T_\la$ in 
eqs. (\ref{meaneq2}), we obtain the Bogoliubov coefficients
relating inertial modes. And from these coefficients,
one gets the mean value of energy flux $T_{VV}$
as in \reff{newTvv}, but with causality built in.

We now study the case of a thick mirror 
with $J$ given by $\Phi^\dagger i \raise 0.1mm 
\hbox{$\stackrel{\leftrightarrow}{\di_\tau}$} \Phi $.
To display the effects of $f(\rho)$,
it is convenient to work with the (spatial) Fourier components.
Eqs. (\ref{phi+in+out}) become 
\ba
\varphi_{\la,l} 
&=& 
\varphi_{\la,l}^{in} 
- \frac{2g\la}{(\la+i\e)^2-l^2} \: \intelp f_{l-l'} \: \varphi_{\la,l'} 
\nn
&=& 
\varphi_{\la,l}^{out} 
- \frac{2g\la}{(\la-i\e)^2-l^2} \: \intelp f_{l-l'} \: \varphi_{\la,l'}\ .
\ea
For an arbitrary window function $f$, 
these equations do not lead to analytic relations
between asymptotic $in$ and $out$ fields. 
Therefore, we use perturbation theory.
To first order in $g$ we get
\ba
\T^f_\la = - g \:
\left(
\begin{array}{cc}
 1 &  2 \pi \: f_{2\la} \\
\disp  2 \pi \: f^*_{2\la} &  1 
\end{array}
\right)
\ea  
For a normalized gaussian function $f$ centered on $\rho_0$, 
the non-diagonal terms which determine the reflection 
probability are 
$g e^{\pm 2i\la\rho_0} e^{-2\la^2\sigma^2}$.
Therefore, $\sigma$, the spread of the mirror 
reduces the reflection of high frequencies: 
for $\la \gg 1/\sigma$ the mirror is completely transparent
(this is also true for the two other $J$'s).

\subsection{The scattering with $g$ time dependent}
\par

In this subsection,
the coupling parameter is a function of the proper time 
$g(\tau) = g \: f(\tau)$ where $f(\tau)$ is normalized by
$\intetau f(\tau) = 2T$ with $2T$ the proper-time lapse
during which the interactions are turned on. 
Unlike what we had in the former subsection, 
resumming the Born series is no longer possible 
since the time dependence of the coupling 
destroys the decoupling of the equations
into sectors at fixed frequency $\la$.
In fact we meet a situation analogous of that of a thick mirror 
which mixed different momenta.
Thus, we shall work perturbatively:
all quantities will be
evaluated up to the second order in $g$.

We remind the reader that in the interacting picture, 
the operator $\Phi$ evolves freely, \ie with $g=0$.
Therefore the $in$ 
operators $a_\om, b_\om$ coincide with the $out$ operators
and are equal to the usual Minkowski operators.
Hence, they define the (Minkowski) vacuum $\vac$. 
Instead,  the states evolve through the action of 
the time ordered operator:
\be
\ket{\Psi(t=+\infty)} = T \: e^{\disp i  L} \, \ket{\Psi(t=-\infty)}\, ,
\ee
where $L=g \int d\tau f(\tau) J$. 
Since the trajectory is timelike
the time ordering with respect to the Minkowski time $t$
is equivalent to that of the proper-time $\t$.

To make contact with Section 2,
we shall work in this section with the state
$\ket{\Psi_0(t)}$ which is equal to 
$\vac$ for $t=\t=-\infty$.  
When expressing its final value in the basis of the 
unperturbed states, \ie the states which would
have been stationary in the absence of interactions,
we get
\ba \label{state} 
\ket{\Psi_0(\t=+\infty)} 
&=& \vac  + 
\sum_{i,j}\int \!\!\!\!\intoom \!d\omp \;
\big( B_{\om \om'}^{i j } 
+   C_{\om \om'}^{i j }
\big)
\; \ket{\om \om' }_{ij} 
\ea
where 
\ba \label{defB}
B_{\om \om'}^{i j } 
&=& ig \: 
\bvac \; 
a_\om^i \: b_{\om'}^j \; 
\Big( \intetau f(\tau) \; J(\tau)
\Big) \; \vac
\, , \quad \ket{\om \om' }_{ij} \equiv a_\om^{i \; \dagger} \: 
b_{\om'}^{j \; \dagger} \vac \, ,
\\
\label{C} 
C_{\om \om'}^{i j } 
&=& - g^2
\bvac \; 
a_\om^i \: b_{\om'}^j \; 
\Big(\disp\intetau \int_{-\infty}^{\disp \t}\!  d\t^\p \;  
f(\t) \; f(\t ') \; J(\t) \; J(\t^\p)\Big ) 
\; \vac_c .
\ea
We have limited the expansion in $g$
to these three terms since we shall
compute the energy-momentum tensor up to $g^2$ terms only. 
As before, $i,j$ denote the $U,V$ sectors
and $\om,\omp$ Minkowski energies.
The symbol $\langle \: \rangle_c$ means that 
only the {\it connected} part of the expectation value is kept. 
This restriction follows from the fact that the contribution
of the disconnected graphs cancels out 
since they also appear in the denominator of the
expectation values, see \eg \cite{IZ}.

Using \reff{state}, the expectation value of $\TVV$ is given by :
\ba \label{TVVintfinal}
\ave{\TVV}
&=& 
\bra{\Psi_0(\t=+\infty)} \TVV \ket{\Psi_0(\t=+\infty)}_c
\nn
&=& 
\re 
\left[ 
\sum_j \int\!\!\!\!\intoom \! d\omp \frac{\sqrt{\om\omp}}{2\pi}
\right.
\left[ 
e^{-iV(\omp-\om)}  \intok 
\Big( 
B_{\om k}^{V j \: *} B_{\omp k}^{V j } + 
\bar B_{\om k}^{V j \: *} \bar B_{\omp k}^{V j }
\Big) \quad \quad
\right.
\nn
&&\quad \quad - 
\left. 
\left.
e^{-iV(\omp+\om)} \; 
\left( 
B_{\om \omp}^{V V} + \bar B_{\om \omp}^{V V} 
+ C_{\om \omp}^{V V} + \bar C_{\om \omp}^{V V}  
\right)
\right]  
\right] \ ,
\ea
where $\bar B_{\om k}^{i j } \; \and \bar C_{\om \omp}^{V V}$ 
are related to the unbar quantities 
by inverting particle and anti-particle operators,
thus $\bar B_{\om \omp}^{i j } = B_{\omp \om}^{ji}$ and 
$\bar C_{\om \omp}^{V V} =   C_{\omp \om}^{V V}$.

Since the integral of the second term in \reff{TVVintfinal} vanishes
and since bar and unbar quantities differ at most by a phase, 
the total energy received on the $V$ part of $\scryp$ is :
\ba \label{HVfinal}
\ave{H_{V}} 
\equiv \inte \! dV \; \ave{\TVV}_c
= 2  \: \sum_j \intoom \om  \intok 
|B_{\om k}^{V j }|^2 
\, .
\ea
Hence only the $B$ terms contribute to the energy 
as the $\bt$ terms did in \reff{H}. 

In order to compute the local properties of the flux,
we need to compute the second term of \reff{TVVintfinal}.
To this end 
we decompose $C_{\om \omp}^{V V }$ into two parts :
\ba \label{C=R+D}
C_{\om \omp}^{VV} 
= R_{\om \omp}^{VV}
- \bvac \; a_\om^V \: b^V_{\omp} \; {\cal D} \; \vac_c \ ,
\ea 
where 
\ba \label{R}
R_{\om \omp}^{VV} &=&  -
\half \: \bvac \; a_\om^V \: b^V_{\omp} \; 
\: L\:  L \: \; \vac_c \ , 
\\
{\cal D} 
&=& {g^2 \over 2}
\Big(
\disp\intetau \intetaup f(\t) \; f(\t ') \; \e(\t-\t') \; J(\t) \; J(\t^\p) 
\Big)
\ea
and
$\e(\t-\t') = \theta(\t-\t')-  \theta(\t'-\t)$. Then, 
$\ave{\TVV}_{\cal D}$,
 the contribution of ${\cal D}$ to $\ave{\TVV}$,
 enjoys the following properties. 
First it carries no energy. This is obvious since it is built 
with terms which all contain $e^{iV(\om + \omp)}$.
Secondly, it 
vanishes for $f(\t) = cst$.
This can be understood from the fact that 
the time ordering properties can be encoded in the
analytical properties of the matrix $\T_\la$ 
which is diagonal in $\la$, see \reff{analT}.
This means that this term modifies the shape of the transients 
related to the switching on and off of the interaction
but without affecting their energy content.
In the rest of the paper,
we shall therefore ignore this term.

We now compute $R_{\om \omp}^{VV}$.
Since only the connected part should be kept,
we can insert the following operator between the two operators 
$L$ in \reff{R}
\ba \label{projector}
\sum_{i,j} \intok \intokp 
{a_k^i}^\dagger \; {b^j_{\kp}}^\dagger 
\vac \bra{0} \; 
{a_k^i} \; {b^j_{\kp}}  \ .
\ea
Grouping together, as in \reff{TVVintfinal}, the first order and the second 
order contribution in $g$, we get
\ba
B_{\om \omp}^{V V } + \bar B_{\om \omp}^{V V } 
+ R_{\om \omp}^{V V } + \bar R_{\om \omp}^{V V }
= \sum_i \intok  
(\bar A_{\om k}^{V j \: *} \; B_{\omp k}^{V j}  
+ A_{\omp k}^{V j \: *}  \; \bar B_{\om k}^{V j }) 
\ea
with 
\ba \label{defA}
A_{\om k}^{i j \: *} = \bvac\; a_\om^i \; 
\Big( 1 + i L \Big)  \;  
{a_k^j}^\dagger \; \vac_c
\; \; \and \; 
\bar A_{\om k}^{i j \: *} 
= \bvac \; b_\om^i \; 
\Big( 1+iL \Big) \;  
{b_k^j}^\dagger \; \vac_c \ .
\ea
Hence we find that $\ave{T_{VV}}$ is given by \reff{newTvv0}
with
\ba 
\label{Tvvtd}
\ave{\TVV}^{particle}
&=&  \re \sum_j \intoom \intoomp \frac{\sqrt{\om \omp}}{2\pi}
\\
&&
\Big[e^{-iV(\omp-\om)} \; 
\left(
\intok {B_{\om k}^{V j }}^* B_{\omp k}^{V j } 
\right)
-
e^{-iV(\omp+\om)} \; 
\left(
\intok  \bar A_{\om k}^{V j \: *} \; B_{\omp k }^{V j} 
\right) \Big]\ . \nonumber
\ea
$\ave{\TVV}_{anti-particle}$ is given by
the same expression 
with $\bar A_{\om k}^{V j },B_{\om k}^{V j }$ 
replaced by $A_{\om k}^{V j },\bar B_{\om k}^{V j }$.

Thus, to second order in $g$, 
we recover the structure of \reff{newTvv}
which gives the flux emitted by a partially transmitting mirror.
The Bogoliubov coefficients $\al_{\om k}^{V j }$ and 
${\beta_{\om k}^{V j }}$ have been replaced by the transition amplitudes 
$A_{\om k}^{V j }$ and ${B_{\om k}^{V j }}$.
In this we recover the correspondence of eqs. (\ref{physa}) 
and (\ref{physb}) when considered to first order in the transfer
matrix $\T_\la$. 
This is not surprising since the evolution operator $Te^{iL}$
which defines $A_{\om k}^*$ and ${B_{\om k}}$, given in eqs. 
(\ref{defA}) and  (\ref{defB}),
{\it is} the operator $\Ss$ of \reff{Sso}.

This correspondence is nicely illustrated in the case where $g(\tau)=g$
and $J=\Phi^\dagger i \raise 0.05mm 
\hbox{$\stackrel{\leftrightarrow}{\di_\tau}$} \Phi$. 
In this case, to order $g$
but whatever is the mirror's trajectory $U=U_{cl}(V)$,
one has the following identities
\ba
A_{\om k }^{V U} = { g }\, \al_{\om k }\, , 
\quad 
B_{\om k }^{V U} = { g }\, \bt_{\om k }\, ,
\ea
where $\al_{\om k }$ and $\bt_{\om k }$
are the Bogoliubov computed in the DF model. 
These relations establish that $\al_{\om k }$ and 
$\bt_{\om k }$ should be understood as transition amplitudes.
This implies in particular that the momentum transfers to
mirror (which have been neglected so far) 
are respectively $\hbar (k+\om)$ and $\hbar (-k+\om)$. 
This imposes limitations
when considering ultra-high (trans-Planckian) frequencies 
since neglecting the momentum transfers requires $\hbar \om 
\ll M$ where $M$ is the mass of the mirror\cite{recmir}.
Thus, when high frequency quanta are emitted, 
the validity of the predictions obtained with 
a recoil-less model {\it must} be 
questioned\cite{P:hawkfeynman}.

\section{Applications}
\par

First, we analyze the properties of the transients 
associated with the switching on and off
when the mirror is at rest ($z=0$) and in Minkowski vacuum. 
Secondly, we generalize this analysis by replacing the 
Minkowski vacuum by a thermal bath. 
Then, we use the well-known parallel between inertial systems 
in a thermal bath and uniformly accelerated systems in vacuum
to obtain a regularized expression of
the flux emitted by a uniformly accelerated mirror.

\subsection{The transients in vacuum}

\par

We first focus on the frequency content
of the transients.
For an inertial mirror at rest at $z=0$ in Minkowski vacuum,
the transition amplitudes 
$A$ and $B$ of eqs. (\ref{defA}) and (\ref{defB}) 
can be expressed in terms of the Fourier transforms of $f(t)$ 
\be
f_{\om} = \frac{1}{2\pi} \int \! dt \; f(t) e^{i \om t} \ .
\label{fourierf}
\ee
To order $g$ we obtain
\ba
\label{fourierAB}
A_{\om k}^{i j \: *} 
&=& \delta(\om - k) \delta^{i j } + i g  \: f_{\om - k} \:
 \frac{j(\om,k)}{\sqrt{\om k}} \ ,
\\
B_{\om k}^{i j } 
&=& i g \: f_{\om + k} \: \frac{j(\om, -k)}{\sqrt{\om k}}\ ,
\ea
where
\ba
\label{3j}
j(\om,k) = 
\left\{
\begin{array}{l}
1 \; \;  \;  \quad \quad \,
\for \Phi^\dagger \Phi + \Phi \Phi^\dagger \\
\om + k  \; \; \; \;  \for \Phi^\dagger i \raise 0.05mm  
\hbox{$\stackrel{\leftrightarrow}{\di_t}$} \Phi \\
\om k  \;\quad \quad
\for \di_t \Phi^\dagger \di_t \Phi +  \di_t \Phi \di_t \Phi^\dagger
\end{array}
\right. \ .
\ea
Thus to order $g^2$, the mean number of $V$ particles 
of energy $\om$ is given by :
\ba \label{Nomega}
\ave{N_\om^V}
=  2 \: \sum_j \; \intok 
|B_{\om k}|^2 \ .
\ea
The factor of $2$ arises from the fact that 
the pair production amplitudes $B$ are independent
of the (relative) sign of the momentum of the two quanta.

Our aim is to describe the transients
associated with the switching on and off of the coupling to the
mirror. To this end and to be concrete, we shall work with 
the function 
\be \label{deff}
f(t) = \half \Big( \tanh(\frac{t+T}{\Delta}) - 
\tanh(\frac{t-T}{\Delta}) \Big) \ .
\ee
It is constant during a lapse of time $2T$ centered about $t=0$ 
and the time intervals of the switching on and off are 
$\simeq 4\Delta$. In the limit $\Delta \to 0$, $f$ tends to the
square window $[\theta(t+T) - \theta(t-T)]/2$. 
The Fourier components of $f$ are  
\be
f_\om = \frac{\Delta}{2} \: \frac{\sin(\om T)}
{\sinh(\om \pi \Delta /2)} \ .
\ee
One sees that the UV behavior is exponentially damped 
by $\Delta$. On the contrary, in the IR, $f_\om \to T/\pi$
since the coupling lasts during $2T$.

When considering the first two cases of 
$j(\om, k)$ of \reff{3j},
these simple observations imply that the mean number $\ave{N_\om}$
is ill defined since the integral over $k$ in \reff{Nomega}
diverges in the IR. Therefore
to obtain well defined expressions we shall work with the 
third case.
In this case one has
\be
\label{Nomega2}
\ave{N_\om^V} 
= \frac{g^2 \Delta^2}{2} \: \om 
\intoomp \omp 
\frac{\sin^2((\om + \omp)T)}
{\sinh^2((\om + \omp)\pi \Delta /2)} \ .
\ee

It is perhaps appropriate to discuss the 
condition on the (dimensionfull) coupling constant $g$ 
which guarantees the validity of a perturbative treatment 
limited to order $g^2$. 
The condition is that the mean number of quanta per quantum cell 
(which is equal to $\ave{N_\om} d\om \simeq \ave{N_\om} \pi / T$ in the limit
$\om T \gg 1$) be well approximated by \reff{Nomega}. 
This requires that the probability to obtain
two quanta in a cell is negligible with respect to that 
to obtain one.  This translates mathematically by
$g^2 \ll T\, \Delta$ in the limit of interest $T/\Delta \gg 1$,
i.e. when the flat plateau is much longer that the 
slopes. The condition $g^2 \ll T\, \Delta$ means
that the limit $T \to \infty$ can be safely taken.
Instead the limit $\Delta \to 0$ is more delicate.
A sufficient condition consists in working 
at fixed $\tilde g^2 \ll 1$ where 
$\tilde g = g (T \Delta)^{-1/2}$.
A stronger condition is to impose that 
the total number of particle emitted,
$\int_0^{\infty} \ave{N_\om^V} d\om$, is finite in the 
limit $\Delta \to 0$. In this case, $\bar g =g/\Delta$
should be held fixed.

When studying \reff{Nomega2},
one first notices that in the limit $T \to \infty$ with $g$ and
$\Delta$ fixed the total number of particles emitted
is independent of $T$, thereby not giving 
rise to a Golden Rule behavior characterized by 
a linear growth in $T$. 
Secondly, $\ave{N_\om^V}$ 
is maximum for $\om \propto 1/\Delta$.
Finally, for  $\om \Delta \gg 1$
one has $\ave{N_\om^V} \simeq e^{-\pi \om \Delta}$. 
We thus find all the expected attributes of transients:
their particle content is independent of the duration $T$
when $T/\Delta \gg 1$, and their Fourier content is peaked
around the adiabatic switching rate $\Delta^{-1}$. 
		
We now study the space time repartition of the energy fluxes
associated with these transients effects. 
We first notice that 
once the \dterm defined in \reff{C=R+D} as been dropped,
the mean flux can be expressed as :
\ba \label{TVVcommu}
\ave{\TVV}
=
- 2 \im \left( \vacave{\TVV L} \right)
+\re \left( \vacave{L\left[ \TVV,L \right]}\right) \ .
\ea
Of course, by decomposing $L$ and $\TVV$ 
in terms of creation and annihilation operators,
one would recover respectively the linear and the 
quadratic contribution of \reff{Tvvtd}. However, being 
interested in the space time properties, we shall not do so
and shall work instead in the time `representation' with
the $V$ part of 
the (positive frequency) Wightman function. It obeys 
\be \label{W0}
\di_{V} W(V-V') = \di_{V} \expect{0}{\Phi^\dagger(V,U) \: \Phi(V',U')}{0} 
= - \frac{1}{4\pi} \: \frac{1} {V-V'-i\e}. 
\ee
Using this function, the linear contribution in $g$ 
reads
\ba \label{TVVlocal1}
\ave{\TVV}_{lin}
&=& - 8 g \: \im \left[ 
\int \! dt \; f(t) \: \left\{ 
\di_{t} \di_V W(V-t) \right\}^2 \right]
\nn
&=&  \frac{g}{12 \pi} \: \di_{t}^3 f(t=V) \ .
\ea
To obtain this result, we have integrated by part three times.
The boundary contributions all vanish since $f$
given in \reff{deff} decreases faster that any power of $t$.
The last integration is trivially performed by using
$ \im ((x -i\e)^{-1}) = \pi \: \delta(x)$. These properties explain
the local character of the expectation value\footnote{
It should be pointed out that we could have written
$\ave{\TVV}_{lin}$ as a commutator. This however is not appropriate
since one looses the analytical properties 
of $W$ which are encoded by $i\e$ (They arise from
frequency content of the vacuum and play a crucial
role in defining the above expressions). 
By performing first the commutator 
(or equivalently by first taking the imaginary part in
\reff{TVVlocal1}) 
one would obtain an ill defined expression. 
The same remark
applies to the quadratic term in $g$. 
To obtain well defined expressions,
only one commutation (and not two) should be done.}. 

To evaluate the quadratic contribution in $g$ we proceed
along the same lines. We first evaluate the commutator 
so as to obtain a quadratic form in $\Phi$ and $\Phi_V$,
where $\Phi_V$ means only the $V$ part of the field operator
$\Phi$ should be kept. We notice that the derivatives $\partial_t$ 
in $J$ might be expressed as $\di_V$ since they are evaluated at $z=0$
but they act both on the $V$ and the $U$ part of $\Phi$. 
Using this notation, one finds
\ba
[\TVV,L]
&=& 
i g \: f(V) \: 
\left[\left(
\di_V \Phi_V^\dagger \: \di^2_V \Phi
+ \di_V \Phi_V \: \di^2_V \Phi^\dagger 
\right) + (\rm{h.c}) \right]
\nn
&&+ 
i g \: f'(V) \: 
\left[ \left( 
\di_V \Phi_V^\dagger \: \di_V \Phi
+ \di_V \Phi_V \: \di_V \Phi^\dagger 
\right) + (\rm{h.c}) \right]  \ . 
\ea
Then the $g^2$ contribution of $\TVV$ is
\ba \label{TVVlocal2}
\ave{\TVV}_{quadr} &=& 
16 g^2 \: f(V) \:
\re 
\left( i 
\int \! dV' \; f(V') \: 
\left( \di_{V'} \di^2_{V} W(V'-V) \right)  
\left( \di_{V'} \di_{V} W(V'-V) \right)
\right)
\nn
&&+ 16 g^2 \:  \di_Vf \:
\re 
\left( i 
\int \! dV' \; f(V') \: 
\left( \di_{V'}  \di_{V} W(V'-V) \right)^2 
\right)
\nn
&=& 
- \frac{g^2}{12 \pi} \: 
\left( f \: \di_V^4 f + 2 \: \di_V f \:  \di_V^3 f \right)
\ .
\ea
Since neither $f$ appears in \reff{TVVlocal1} 
nor $f^2$ in \reff{TVVlocal2},
one recovers the fact that an inertial mirror 
doesn't radiate while its coupling is constant.
This is illustrated in Fig. 1.
\begin{figure}[ht] \label{Tvvfig}
\epsfxsize=6.5cm
\centerline{\rotatebox{-90}{\epsfbox{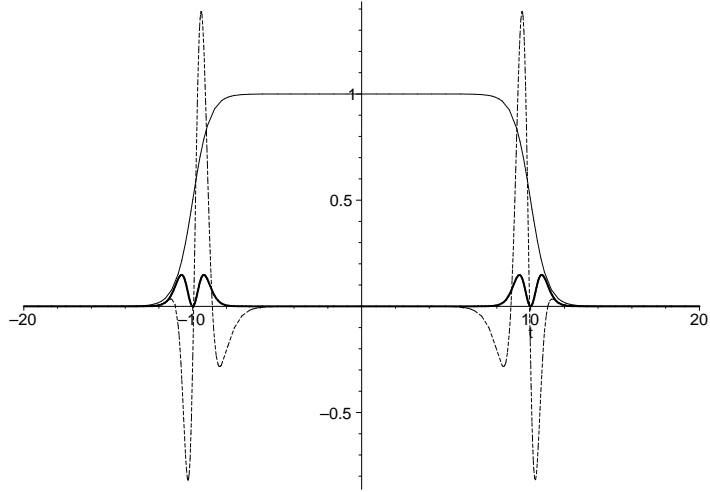}}}
\caption{The thin solid line is $f(t)$ given by \reff{deff},
for $T=10$ and $\Delta=1$.
The dashed line is $\ave{\TVV}_{quadr}$ and the thick line is
the part of $\ave{\TVV}$
which contributes to the energy, see \reff{Hlocal}. 
These two curves have been plotted
in the same arbitrary units.
The behavior of $\ave{\TVV}_{lin}$ is similar
to $\ave{\TVV}_{quadr}$.}
\end{figure}

As in \reff{TVVDFlocal}, one can decompose $\ave{\TVV}$ into two parts : 
a positive definite term and a total
derivative which does not
contribute to the total energy
\ba
\label{Tvvlocal4}
\ave{\TVV}
&=& \ave{\TVV}_{lin} + \ave{\TVV}_{quadr} 
\nn
&=&  \frac{g^2}{12\pi} \Big(\di_V^2 f \Big)^2
- \frac{1}{12\pi} 
\di_V \left[
-g \: \di_V^{2}f + g^2 \: ( \half \di_V^{4}(f^2) - \di_V^{2}((\di_Vf)^2) )
\right]\, .
\ea
Thus the total energy is 
\ba \label{Hlocal}
\ave{H_V} = \frac{g^2}{12\pi} \intev \Big(\di_V^2 f \Big)^2
=\intoom \om \ave{N_\om^V} \,.
\ea
$\ave{H_V}$ is finite when the mean number $\ave{N_\om}$ 
decreases faster than $\om^{-2}$. This is the case
when working with \reff{deff} at fixed $\Delta \neq 0$.
In this case, 
one finds
\ba \label{HF}
\ave{H_V} = \frac{2 \, g^2}{45 \pi} \; 
\frac{1}{\Delta^3} \; F_1(\frac{T}{\Delta}) \ .
\ea
The main feature $\ave{H_V}$
is that it is independent of $T$ in the limit 
$T/\Delta \gg 1$, see Fig. 2. 
\begin{figure}[ht] \label{Hfig}
\epsfxsize=6.5cm
\centerline{\rotatebox{-90}{\epsfbox{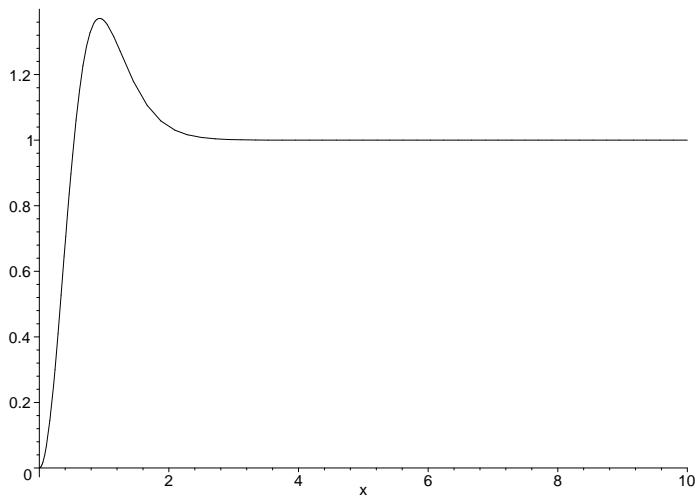}}}
\caption{The plot $F_1(x)$ defined in \reff{HF} in such a way that 
$F \to 1$ for $x\to \infty$ where $x=T/\Delta$.}
\end{figure}

In conclusion, we consider the limit $\Delta \to 0$ which
corresponds to the situation studied in \cite{AndersonDeWitt} and
in \cite{Strominger} in view of its analogies with the 
residual flux emitted at the end of the evaporation of a black hole.
In this limit, $f(t)$ becomes a step function, 
the energy flux is concentrated in a narrow
lapse $\Delta$ and its frequency content diverges. 
In fact $\ave{T_{VV}}$ becomes a distribution since it is built
on the derivatives of $f(t)$. 
However the singularity is worse than a delta,
as clearly seen from \reff{TVVlocal2}. This means that the total
energy emitted is also singular, as indicated in \reff{HF}. 
Moreover, there is no 
reason for the nature of the singularity to be universal.
It depends on the number of derivatives in the Hamiltonian
and it might also vary when considering higher orders in $g$.  
Hence the question: what is the flux emitted by the disappearance
of the reflection condition, is not univocally defined. 
To have a well-defined question, one should first choose
a regular model such as that defined by \reff{lagrangian} with 
$J$ containing at least two derivatives and $g(t)$ given by
\reff{deff}, and only then consider the singular limit.


\subsection{Flux and energy in a thermal bath}

\par
Our aim is to obtain a regularized expression
for the flux emitted by a uniformly accelerated mirror. 
To this end, we shall use the isomorphism 
between the flux emitted by a mirror 
at rest in a heat bath at temperature $a/2\pi$ 
and the Rindler flux emitted by a uniformly accelerated mirror 
of acceleration $a$. 

In a thermal bath, the $V$ part of the Wightman function obeys
\be \label{Wthermal}
\di_V W^\bt(V-V')
= - \frac{1}{4\pi} \frac{\pi}{\bt} \: 
\coth \left(\frac{\pi}{\bt}(V-V'-i\e)\right) \ .
\ee
It reduces to $\di_V W$ of \reff{W0} in the zero temperature limit, i.e. 
for $\bt \rightarrow \infty$.
When replacing $W$ by $ W^\bt$ in eqs. (\ref{TVVlocal1}) and 
(\ref{TVVlocal2}) we obtain the mean flux emitted in a thermal bath.
It can be shown to be\footnote{The details of the 
calculation will be presented in \cite{next}.}
\ba \label{TVVthermallocal}
\ave{\TVV}^\bt
&=&  
\frac{g}{12\pi} 
\di_V^3 f - 
\frac{g^2}{12 \pi}
\left(
f(V) \: \di_V^4 f + 2 \di_V f \: \di_V^3 f \right) 
\nn
&& 
- \Big( {2 \pi \over \bt }\Big)^2   \, \Big[ \frac{g}{12\pi} \di_V f -
\frac{g^2}{12 \pi}
\left(
f(V) \: \di_V^2 f + 2 \di_V f \: \di_V f
\right)
\Big] \ .
\ea
The first two terms are equal to \reff{Tvvlocal4}
and the last two scale like $(\Delta/\bt)^2$.
Thus, they are negligible in a low temperature limit, 
$\beta \gg \Delta$, and dominant in the high temperature regime.

We are now in position to obtain a regular expression for the
flux emitted by a uniformly accelerated mirror in Minkowski
vacuum. Using the well-known isomorphism between systems
at rest in a thermal bath and accelerated systems in vacuum, 
the mean flux of Rindler energy emitted by a mirror
of acceleration $a$ is
\be
\label{Tvvacc}
\ave{T_{vv}(v)}^{acc} = \ave{T_{VV}(V=v)}^{\bt = 2\pi /a} \, , 
\ee
where $v$ is the null advanced Rindler time ($av= \ln(aV)$)
when the mirror is located in the Right Rindler quadrant,
 ($z < \vert t \vert)$. 
The coupling between the mirror and the field is turned on
during a proper time lapse $2T$ and the switching on and off rate 
$\Delta^{-1}$ is also measured with the  proper time.

In the limit $T \gg \bt\,  \and a^{-1}$, $\ave{T_{vv}(v)}^{acc} \to 0$
at fixed $\vert v\vert < T$ since the flux is localized in the 
transients of `thickness' $\Delta$. In this we recover the 
well known result that a uniformly accelerated mirror does not
radiate. In the DF model, this immediately follows
from \reff{TVVDFlocal}. (Notice that this vanishing is a universal
property of accelerated systems when they have reached equilibrium
with the Rindler bath\cite{grove2,RSG,Primer,MaPa}.)
However this vanishing flux is accompanied by transients
effects whose Minkowski properties become singular 
in the limit $T \to \infty$ whatever is the value of $\Delta$. 
To have regular Minkowski properties means that the total energy 
\be
\label{Hacc}
\ave{H_V^{acc}} = \intevv e^{av} \: \ave{T_{vv}^{acc}} \ ,
\ee
and the mean number of Minkowski quanta emitted by the mirror
be finite.

When requiring that $\ave{H_V^{acc}}$ is finite, 
the coupling $f(\tau)$ must decrease faster than the 
Doppler effect $dV/dv = e^{a v}$. 
Using \reff{deff}, this implies $a \Delta <  2$. 
(A similar condition also applies when considering 
the fluxes emitted by an accelerated two level atom\cite{MaPa}.)
Using eqs. (\ref{TVVthermallocal}), (\ref{Tvvacc}) and (\ref{Hacc}),
we get
\be
\label{Hacc2}
\ave{H_V^{acc}} =  \frac{g^2}{12\pi} \intevv e^{av} \: 
\Big[  \Big(\di_v^2 f \Big)^2 + 2 a^2  \Big(\di_v f \Big)^2 \Big] \, .
\ee
What we have learned from the condition $a \Delta <  2$ 
is that the Rindler fluxes which lead to finite Minkowski 
energies are dominated by the vacuum effects governed by 
$\Delta $ rather than by the temperature effects 
induced by the acceleration. 

\begin{appendix}
\section{The $in-out$ overlap in the non-stationary case}
\indent 
In order to have simple expressions for this 
overlap, we will use a discretized basis of wave packets
in which the integrals are replaced by sums 
and Dirac distributions by Kronecker symbols.

Instead of working with the $in$ and $out$ basis,
it is appropriate to define a third class
of operators $\a_\om, \b_\om$. This new basis
generalizes the Unruh modes\cite{Unruh,Primer} 
since $\a_\om$ ($\b_\om$) is made out of $a^{in}_k$ ($b^{in}_k$) 
but is characterized by a fixed $out$ frequency $\om$:
\ba
\al_{\om} \: \a_\om 
&=& \sum_k \al_{\om k}^* \: a_k^{in} \, , 
\,\,\quad \al_{\om} \: \b_\om 
= \sum_k \al_{\om k}^* \: b_k^{in} \, .
\ea
The real coefficients $\al_{\om}$ are such that 
$[\a_\om,\a_\om^\dagger] = 1$,
therefore 
$\al_\om^2 = \sum_k | \al_{\om k} | ^2$. 
The notion of particle/anti-particle is obviously maintained
since the $\a$ are made of $a^{in}$ only.
Notwithstanding, for arbitrary $\al_{\om k}$ and $\bt_{\om k}$
this new basis is not orthogonal 
and the commutation rules 
are given by
\ba \label{commutationrules}
[\a_\om,\a^\dagger_{\omp}]
\equiv  F_{\om\omp}
= [\b_\om,\b^\dagger_{\omp}] = 
{ \disp \sum_k \al_{\om k}^* \al_{\omp k} 
\over \al_{\om} \al_{\omp}}
\ . 
\ea
By construction from \reff{bogop1},
these new operators are related the $out$ operators by 
\ba \label{bogop2}
\left\{
\begin{array}{rc} 
\disp a^{out}_\om  
=  
\al_{\om} \: \a_\om - \sum_{\omp} \al_{\omp} B_{\om \omp}\: \b_{\omp}^\dagger
\\
\disp b^{out}_\om  
=  
\al_{\om} \: \b_\om - \sum_{\omp} \al_{\omp} B_{\om \omp}\: \a_{\omp}^\dagger
\end{array} 
\right. 
\with B_{\om \omp}
&\equiv& \sum_k \bt_{\om k} \: \al^{-1}_{k \omp} \ ,
\ea
where $\al^{-1}_{k \om}$ 
is the inverse matrix of $\al_{\om k}$.
( $\al_{\om k}$ is always invertible 
since otherwise 
there would exist incoming particles
whose scattering would give only anti-particles.)

As for the Unruh modes, the new basis
is useful to relate in a simple way 
the $out$ vacuum to the $in$ vacuum.
Straightforward algebra indeed gives 
\ba \label{vacvac}
\ket{0_{out}} 
= \frac{1}{Z} \:
\exp \Big( \;
{\disp
\sum_{\om \omp k}
\frac{\al_{\omp}}{\al_k} \:
F^{-1}_{\om k} \:
B_{k \omp} \:
\a^\dagger_{\om}
\b^\dagger_{\omp}
\Big)
}
\; \ket{0_{in}} \ ,
\ea
where $Z$ is defined by
\be \label{defZ}
Z^{-2}
 = \vert \scal{0_{out}}{0_{in}} \vert ^2 \ .
\ee

Even though \reff{vacvac} looks cumbersome,
one easily verifies that, to order $\bt^2$, 
it correctly gives the relationship between 
the vacuum decay ($Z>1$) and 
the pair creation probability of 
Minkowski quanta. 
Indeed, 
using $(F^{-1})_{\om \omp} \expect{0_{in}}
{\a_{\omp} \: \a^\dagger_{\om''}}
{0_{in}}= \delta_{\om,\om''}$ and the condition on $B_{\om \omp}$
and $F_{\om \omp}$ which arises from $[a^{out}_\om, b^{out}_{\omp}]=0$
and \reff{commutationrules},
one obtains, 
\be \label{vacdecay}
Z^2 
= 1 +  \sum_{\om \omp} \: 
\Big\vert 
B_{\om \omp}
\Big\vert ^2 + O(\bt^4)\ .
\ee
This is the correct expression since the 
probability to have a pair of $out$ quanta
is
\be
\vert \expect{0_{out}}
{a_\om^{out} \: b_{\omp}^{out}}
{0_{in}} \vert^2  = \Big \vert \frac{B_{\om \omp} }{Z} \Big \vert^2
= \vert {B_{\om \omp}} \vert^2 + O(\bt^4)
\ee

For completeness, we notice that
when the scattering is stationary (as it is the case 
for uniform acceleration and in black hole evaporation), one has 
\be
B_{\om \omp} = \frac{\bt_\om}{\al_\om} \: \delta_{\om,\omp}\, ,\,
F_{\om \omp} =  \delta_{\om,\omp} \, .
\ee
Since they are diagonal, \reff{vacvac} becomes
\be 
\ket{0_{out}} = \frac{1}{Z} \:
\exp \Big( \;
{\disp
\sum_{\om}
\frac{\bt_\om}{\al_\om} \:
\a^\dagger_\om
\b^\dagger_\om
\Big)
}\; \ket{0_{in}} 
\ee
thereby recovering the usual diagonal
expression governed by the ``Unruh'' operators $\a_\om, \b_\om$.

\end{appendix}


\end{document}